\documentclass[aip,jcp,reprint]{revtex4-1}

\usepackage{physics,float,xcolor,amsfonts,amssymb,amsmath,graphicx}
\usepackage{hyperref}
\usepackage[version=4]{mhchem}

\newcommand{\Eh}{$E_\text{h}$}
\newcommand{\mEh}{m\Eh}
\newcommand{\alert}[1]{#1}
\newcommand{\mc}[1]{\multicolumn{1}{c}{#1}}
\newcommand{\Ndet}{N_\text{det}}
\newcommand{\Nbas}{N_\text{bas}}
\newcommand{\Nel}{N_\text{el}}
\newcommand{\tCPU}{t}
\newcommand{\hH}{\Hat{H}}
\newcommand{\hV}{\Hat{V}}

\newcommand{\mA}{\mathcal{A}}

\newcommand{\mD}{\mathcal{D}}

\newcommand{\mU}{\mathcal{U}}
\newcommand{\Ezero}{E^{(0)}}
\newcommand{\Etwo}{E^{(2)}}
\newcommand{\EtwoS}{E_\text{S}^{(2)}}
\newcommand{\EtwoD}{E_\text{D}^{(2)}}

\newcommand{\mDD}{\mD_\text{D}}

\begin{document}
\title{Hybrid stochastic-deterministic calculation of the second-order perturbative contribution 
of multireference perturbation theory}

\author{Yann Garniron}
\affiliation{Laboratoire de Chimie et Physique Quantiques, Universit\'e de Toulouse, CNRS, UPS, France}
\author{Anthony Scemama}
\thanks{Corresponding author}
\email{scemama@irsamc.ups-tlse.fr}
\affiliation{Laboratoire de Chimie et Physique Quantiques, Universit\'e de Toulouse, CNRS, UPS, France}
\author{Pierre-Fran\c{c}ois Loos}
\affiliation{Laboratoire de Chimie et Physique Quantiques, Universit\'e de Toulouse, CNRS, UPS, France}
\author{Michel Caffarel}
\affiliation{Laboratoire de Chimie et Physique Quantiques, Universit\'e de Toulouse, CNRS, UPS, France}

\begin{abstract}
A hybrid stochastic-deterministic approach for computing the second-order perturbative contribution $\Etwo$ within multireference perturbation theory (MRPT) is presented. 
The idea at the heart of our hybrid scheme --- based on a reformulation 
of $\Etwo$ as a sum of elementary contributions associated with each determinant of the MR wave function --- is to split $\Etwo$ into a stochastic and a deterministic part. 
During the simulation, the stochastic part is gradually reduced by dynamically increasing the deterministic part until one reaches the desired accuracy.
In sharp contrast with a purely stochastic MC scheme where the error decreases indefinitely as 
\alert{$\tCPU^{-1/2}$} (where $\tCPU$ is the computational time), the statistical error in our hybrid 
algorithm \alert{ displays a polynomial decay $\sim \tCPU^{-n}$ with 
$n=3-4$ in the examples considered here}.
If desired, the calculation can be carried on until the stochastic part entirely vanishes. 
In that case, the exact result is obtained with no error bar 
and no noticeable computational overhead compared to the fully-deterministic calculation.
The method is illustrated on the \ce{F2} and \ce{Cr2} molecules. 
Even for the largest case corresponding to the \ce{Cr2} molecule treated with the cc-pVQZ basis set, very accurate results are obtained for $\Etwo$ for an active space of (28e,176o) and a MR wave function including up to $2 \times 10^7$ determinants.
\end{abstract}
\maketitle

%%%%%%%%%%%%%%%%%
\section{Introduction}
%%%%%%%%%%%%%%%%%
Multireference (MR) approaches are based upon the distinction between
non-dynamical (or static) and dynamical correlation effects.  Though such a
clear-cut distinction is questionable, it is convenient to discriminate between
the so-called static correlation effects emerging whenever the description of
the molecular system using a single configuration breaks down (excited-states,
transition-metal compounds, systems far from their equilibrium geometry,
\ldots)\cite{Szalay_2012} and the dynamical correlation effects resulting from
the short-range part of the electron-electron repulsion.\cite{Hattig_2012}

To quantitatively establish this distinction, the Hamiltonian is decomposed as
\begin{equation}
	\hH = \hH^{(0)} + \hV,
\end{equation}
where the zeroth-order Hamiltonian $\hH^{(0)}$ is chosen in conjunction with some MR wave function including the most chemically-relevant configurations at the origin of static correlation effects, and
\begin{equation}
	\hV = \hH - \hH^{(0)}
\label{decomposition}
\end{equation}
is the residual part describing the bulk of dynamical correlation effects. 
The plethora of MR methods found in the literature results from the large freedom in choosing $\hH^{(0)}$, and the fact that $\hV$ may or may not be treated perturbatively.
Among the non-perturbative approaches, let us cite the two most common ones, namely the MR configuration interaction (MRCI)\cite{mrci1, mrci2, mrci3} and the MR coupled cluster (MRCC)\cite{mrcc1, mrcc2, mrcc3, Lyakh_2012} approaches.  
However, because of their high computational cost, these methods are usually limited to systems of moderate size.

To overcome the computational burden associated with these methods --- yet still capturing the main physical effects --- a natural idea is to treat the potential as a perturbation, entering the realm of MR perturbation theories (MRPT). 
Several flavors of MRPT exist depending on the choice of $\hH^{(0)}$
(Epstein-Nesbet decomposition,\cite{epstein, nesbet} Dyall
Hamiltonian,\cite{nevpt1, nevpt2} Fink's partitioning,\cite{fink1, fink2} etc).
Among the most commonly-used approaches, we have the CASPT2\cite{caspt1,
caspt2} and NEVPT2\cite{nevpt1, nevpt2} methods. Regarding the construction of
the zeroth-order part,
CASSCF-type approaches are the most widely-used schemes,\cite{casscf1, casscf2, casscf3} but other methods, such as CASCI, selected CI (see Refs.~\onlinecite{huron_jcp_1973, paci} and references therein), FCIQMC\cite{fciqmc1,fciqmc2,fciqmc3} or DMRG-type approaches\cite{dmrg1, dmrg2, dmrg3} can also be employed. 

In this work, we shall consider MRPT theories limited to the second order in perturbation (MRPT2).\cite{caspt1} 
We address the important problem of calculating efficiently the second-order perturbative contribution $\Etwo$ in situations where standard calculations become challenging.
Here, we suppose that the MR wave function has already been constructed by any method of choice.

\alert{Although the present method can be easily generalized to any externally-decontracted MRPT 
approach (such as the recently-introduced JM-MRPT2 method \cite{Giner_2017}), 
for the sake of simplicity, we shall restrict ourselves here to MR Epstein-Nesbet perturbation theory. 
Extension to externally-contracted methods, such as CASPT2 or NEVPT2, is less obvious --- although not impossible --- since the excited contracted wave
functions are non-orthogonal.}

The computational cost of MRPT2 can rapidly become unbearable when the number of electrons $\Nel$ and the number of one-electron basis functions $\Nbas$ become large.
The cost is indeed proportional to the number of reference determinants $\Ndet$ times the total number of singly- and doubly-excited determinants (scaling as $\Nel^2 \Nbas^2$). 
Because our main goal is to treat large, chemically-relevant systems, the development of fast and accurate schemes for computing $\Etwo$ becomes paramount.
Of course, in actual calculations, a trade-off must be found between the price to pay to build the MR wave function and the effort needed to evaluate $\Etwo$.
Increasing $\Ndet$ (i.e.~improving the MR wave function) may appear as the natural thing to 
do as the magnitude of $\Etwo$ decreases and the contribution of the neglected higher orders 
is made smaller. 
However, its computational price (proportional to $\Ndet$) increases stiffly and the calculation becomes rapidly unfeasible.
Unfortunately, this balance is strongly dependent on the method used to generate the MR wave function and on the ability to compute rapidly and accurately $\Etwo$. 

In this work, we present a simple and efficient Monte Carlo (MC) method for computing the second-order perturbative contribution $\Etwo$. 
For all the systems reported here, the reference space is constructed using the CIPSI method,\cite{huron_jcp_1973, cipsi_1983,paci} a selected CI approach where important determinants are selected perturbatively. 
However, other variants of selected CI approaches or any other method for constructing the reference wave function may, of course, be used.
\alert{Note that, in this study, the reported wall-clock times only refer to the computation of $\Etwo$, i.e.~do not take into account the preliminary calculation of the reference wave function.}

A natural idea to evaluate $\Etwo$ with some targeted accuracy
is to truncate the perturbational sum over excited determinants.
However, since all the terms of the second-order sum have the same (negative) sign, the truncation will inevitably introduce a bias which is difficult to control. 
A way to circumvent this problem is to resort to a stochastic sampling of the various contributions. 
In this case, the systematic bias is removed at the price of introducing a statistical error. 
The key property is that this error can now be controlled thanks to the central-limit theorem.
However, in practice, to make the statistical average converge rapidly and to get statistical error small enough, care has to be taken in the way the statistical estimator is built and how the sampling is performed. 
The purpose of the present work is to propose a practical solution to this problem.

Note that the proposal of computing stochastically perturbative contributions is not new. 
In the context of second-order M{\o}ller-Plesset (MP2) theory, where the reference Hamiltonian reduces to the Hartree-Fock Hamiltonian, Hirata and coworkers have proposed a MC scheme for calculating the MP2 correlation energy.\cite{hirata1, hirata2}
However, we point out that this approach, based on a single-reference wave function, samples a 13-dimensional integral (in time and space) and has no direct relation with the present method. 
In a recent work, Sharma et al.\cite{sharma1} address the very same problem of computing stochastically the second-order perturbative contribution of Epstein-Nesbet MRPT\@.
Similarly to what is proposed here, $\Etwo$ is recast as a sum over contributions associated with each reference determinant and contributions are stochastically sampled.
However, the definition of the quantities to be averaged and the way the sampling is performed are totally different.
Finally, let us mention the recent work of Jeanmairet et al.\cite{sharma2} addressing a similar problem in a different way. 
Within the framework of the recently proposed linear CC MRPT, it is shown that both the zeroth-order and first-order wave functions can be sampled using a generalization of the FCIQMC approach. 
Here also, $\Etwo$ can be expressed as a stochastic average.

The present paper is organized as follows. 
In Sec.~\ref{sec:MRPT2}, we report notations and basic definitions for MRPT2.  
Section~\ref{sec:MC} proposes an original reformulation of the second-order contribution allowing an efficient MC sampling.
The expression of the MC estimator is given and a hybrid stochastic-deterministic approach greatly reducing the statistical fluctuations is presented.
In Sec.~\ref{sec:numerical}, some illustrative applications for the \ce{F2} and \ce{Cr2} molecules are discussed. 
Finally, some concluding remarks are given in Sec.~\ref{sec:conclusion}.

%%%%%%%%%%%%%%%%%
\section{
\label{sec:MRPT2}
Second-order multireference perturbation theory}
%%%%%%%%%%%%%%%%%
\subsection{Second-order energy contribution}
\label{sec:E2}
%%%%%%%%%%%%%%%%%
In MR Epstein-Nesbet perturbation theory, the reference Hamiltonian is chosen to be
\begin{equation}
	\hH^{(0)} = E^{(0)} \dyad{\Psi} + \sum_{\alpha \in \mA} H_{\alpha\alpha}  \dyad{\alpha},
\label{h0}
\end{equation}
where $H_{\alpha\alpha} = \expval{\hH}{\alpha}$ and
\begin{equation}
	\ket{\Psi} = \sum_{I \in \mD} c_I \ket{I}
\end{equation}
 is the reference wave function expressed as a sum of $\Ndet$ determinants belonging to the reference space
\begin{equation}
	\mD = \qty{  \ket{I}, I=1, \ldots, \Ndet },
\end{equation}
and 
\begin{equation}
	E^{(0)} = \frac{\expval{\hH}{\Psi}}{\braket{\Psi}}
\end{equation}
is the corresponding (variational) energy.
The sum in Eq.~\eqref{h0} is over the set of determinants $\ket{\alpha}$ that do not belong to $\mD$ but are connected to $\mD$ via $\hH$: 
\begin{equation}
	\mA = \qty{ \ket{\alpha} \notin \mD \land \qty( \exists \ket{I} \in \mD \mid H_{\alpha I} \ne 0 ) }.
\end{equation}
Due to the two-body character of the interaction, the determinants $\ket{\alpha}$ are either singly- or doubly-excited with respect to (at least) one reference determinant.\cite{SzaboBook}
However, several reference determinants can be connected to the same $\ket{\alpha}$.

Using such notations, the second-order perturbative contribution is written as
\begin{equation}
	\Etwo = \sum_{\alpha \in \mA} \frac{\abs*{\mel{\alpha}{\hH}{\Psi}}^2}{ \Delta E_\alpha},
\label{eq:pt2}
\end{equation}
with $\Delta E_\alpha=E^{(0)}-H_{\alpha\alpha}$.

%%% FIG 1 %%%
\begin{figure}
	\includegraphics[width=0.7\columnwidth]{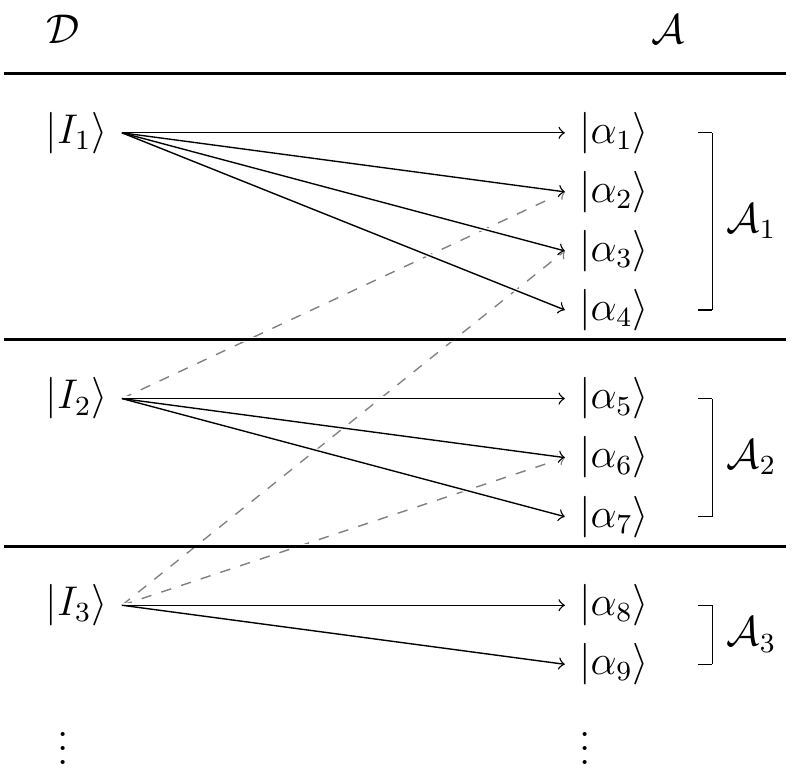}
\caption{
Iterative construction of the subsets ${\cal A}_I$.
Arrows indicate a non-zero matrix element $H_{I\alpha} = \mel{I}{\hH}{\alpha}$. 
Solid arrows: the determinant $\ket{\alpha}$ is accepted as a member of the subset $\mA_I$. 
Dotted arrows: the determinant $\ket{\alpha}$ already belongs to a previous subset $\mA_{J < I}$ and is therefore not incorporated into $\mA_I$.
\label{fig1}}
\end{figure}

%%%%%%%%%%%%%%%%%
\subsection{Partition of $\mA$}
\label{sec:part}
%%%%%%%%%%%%%%%%%
The first step of the method --- instrumental in the MC algorithm efficiency --- is the partition of $\mA$ into $\Ndet$ subsets $\mA_I$ associated with each reference determinant $\ket{I}$:
\begin{equation}
	\mA = \bigcup_{I=1}^{\Ndet} \mA_I 
	\qq{with}
	\mA_I  \cap  \mA_J = \emptyset \qq{if} I \ne J.
\end{equation}
To define ${\cal A}_I$, the determinants $\ket{I}$ are first sorted in descending order according to the weight 
\begin{equation}
	w_I = \frac{c^2_I}{\braket{\Psi}}.
\end{equation} 

The partition of $\mA$ starts with $\mA_1$ defined as the set of determinants
$\ket{\alpha} \in \mA$ connected to the first reference determinant
(i.e.~$I=1$).  Then, $\mA_2$ is constructed as the set of determinants of $\mA$
connected to the determinant corresponding to $I=2$, but not belonging to $\mA_1$.
The process is carried on up to the last determinant.
This partition is schematically illustrated in Fig.~\ref{fig1}. 
Mathematically, it can be written as
\begin{equation}
\label{Partition-A_I}
	\mA_I = 
	\qty{
	\ket{\alpha} \in \mA \mid H_{\alpha I} \ne 0 \land
	(\forall J < I, \ket{\alpha} \notin \mA_J)
	}.
\end{equation}
Because of the way they are constructed, the size of $\mA_I$ is expected to decrease rapidly as a function of $I$, except for a possible transient regime for very small $I$.   

%%% FIG 2 %%%
\begin{figure}
	\includegraphics[width=.95\columnwidth]{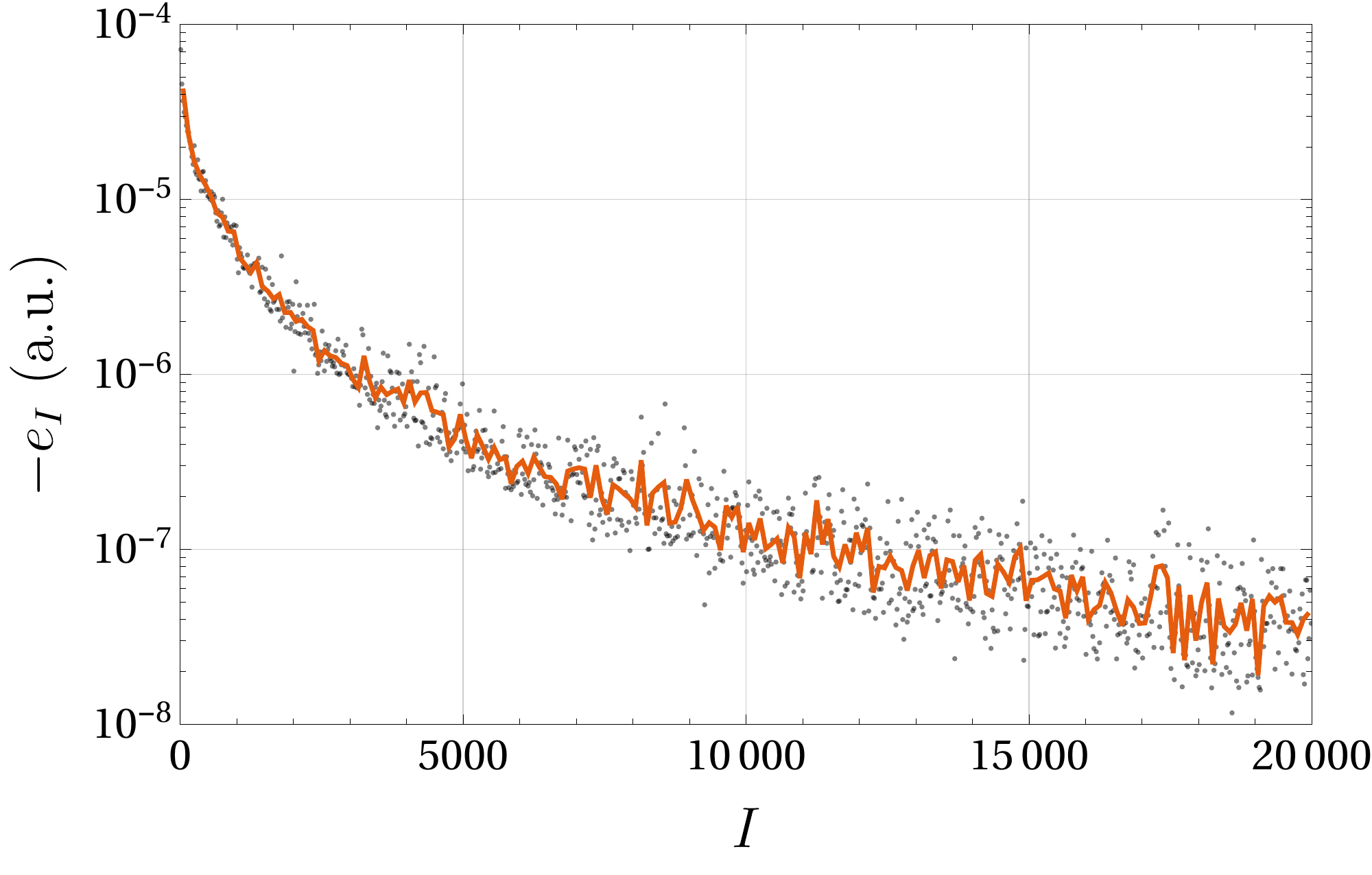}
\caption{$-e_I$ as a function of $I$ for the first $20\,000$ determinants selected by the CIPSI method for the F$_2$ molecule at equilibrium geometry with the cc-pVQZ basis set.
The two sets of data are obtained by averaging either by groups of 20 (point cloud) or 100 (solid line) values.}
\label{fig2}
\end{figure}

\alert{A key point in the construction of the partition of $\mA$ is to avoid both the computation of redundant contributions and the storage of unnecessary intermediates.
First, when a determinant $\ket{\alpha}$ is generated by applying a single or double excitation operator to a reference determinant $\ket{I}$, one has to check that $\ket{\alpha}$ does not belong to $\mD$. 
If the reference determinants are stored in a hash table, 
the presence of $\ket{\alpha}$ in $\mD$ can be checked in constant time.
Next, one has to know if $\ket{\alpha}$ has already been generated via another 
reference determinant $\ket{J}$. To do so, one must compute the number of holes and particles 
between $\ket{\alpha}$ and each determinant preceding $\ket{I}$ in $\mD$. As soon as 
an excitation degree lower than 3 is found, the search can be aborted since the contribution 
is known to have been considered before. In the worst-case scenario, this step scales 
as $\order{\Ndet}$, and the prefactor is very small since finding the excitation degree 
between two determinants can be performed in less than 20 CPU cycles\cite{scemama_2013_3} 
(comparable to a floating-point division). Furthermore, the asymptotic scaling 
can be further reduced by sorting the determinants in groups with same spin string. Indeed, 
one only has to probe determinants $\ket{J}$ that are no more than quadruply 
excited with respect to $\ket{I}$, and, if the search is restricted to groups with spin-up string, 
the asymptotic scaling reduces to $\order{\sqrt{\Ndet}}$.
%Note that, in our implementation, 
%we actually double the storage using both spin-up and spin-down groups to improve the filtering. 
To provide a quantitative illustration of the computational effort associated with the 
construction of the partitioning, using $2 \times 10^7$ determinants (as in the case 
of \ce{Cr2} presented below), this preliminary step is negligible: 
on a single 2.7~GHz core, the calculation takes $20\text{ cycles } 
\times \Ndet^{3/2} / (2.7 \times 10^9\text{ cycles/s}) 
\sim 663\text{ s}$ {(CPU time)}, while the total execution time {(wall-clock time)} of the entire run ranges from $14$~minutes to $18.5$ hr using 800~cores (see Table~\ref{tab1}).}

%%%%%%%%%%%%%%%%%
\subsection{Partition of $\Etwo$}
%%%%%%%%%%%%%%%%%
Thanks to the partition of $\mA$ (see Eq.~\eqref{Partition-A_I}), the sum~\eqref{eq:pt2} can be decomposed into a sum over the reference determinants $\ket{I}$:
\begin{equation}
	\Etwo = \sum_{I=1}^{\Ndet} e_I,
\end{equation}
where
\begin{equation}
	e_I = \sum_{\alpha \in \mA_I} \frac{ \abs*{ \mel{\alpha}{\hH}{\Psi}}^2 }{\Delta E_\alpha}.
\end{equation}
Moreover, noticing that by construction the determinants $\ket{\alpha}$ belonging to $\mA_I$ are not connected to the part of the reference function expanded over the preceding reference determinants, we have
\begin{equation}
\label{e2_contrib}
	e_I =  \sum_{\alpha \in \mA_I} \frac{ \abs*{ \mel{\alpha}{\hH}{\Psi_I}}^2 }{\Delta E_\alpha},
\end{equation}
where
\begin{equation}
	\ket{\Psi_I} =\sum_{J=I}^{\Ndet} c_J \ket{J}
\label{trunc}
\end{equation}
is a truncated reference wave function.
Our final working expression for the
second-order contribution $E^{(2)}$ is thus written as
\begin{equation}
	E^{(2)} = \sum_{I=1}^{\Ndet}  e_I  = \sum_{I=1}^{\Ndet} \sum_{\alpha \in \mA_I} 
	\frac{ \abs*{\mel{\alpha}{\hH}{\Psi_I}}^2 }{\Delta E_\alpha}.
\label{pt2_final}
\end{equation}

A key property at the origin of the efficiency of the MC simulations presented below is that the $e_I$'s take their largest values at very small $I$. 
Then, they decay very rapidly as $I$ increases. 

This important property is illustrated in Fig.~\ref{fig2}. 
The data have been obtained for the \ce{F2} molecule at the equilibrium bond length of $R_\text{F-F}= 1.4119$~\AA\, using Dunning's cc-pVQZ basis set.\cite{Dunning_1989}
The multideterminant reference space is built by selecting determinants using the CIPSI algorithm. 
Figure~\ref{fig2} displays the $e_I$'s for the first 20\,000 selected determinants. 
As one can see, the $e_I$'s decay very rapidly with $I$.
Of course, at the scale of the individual determinant, there is no guarantee of a strictly monotonic decay, and it is indeed what we observe. 
By averaging groups of successive $e_I$'s, the curve can be smoothed out. 
The two data sets presented in Fig.~\ref{fig2} have been obtained by averaging either by groups of 20 (point cloud) or 100 (solid line) values.

It is important to note that the very rapid decay of the $e_I$'s is a direct consequence of the way we have chosen to decompose $\mA$.
To be more precise, we note that in Eq.~\eqref{e2_contrib}, the decay has three different origins:
\begin{itemize}
	\item the number of determinants involved in the sum over $\ket{\alpha}$ decreases as a function of $I$;
	\item the excitation energies $\Delta E_\alpha$ increase with $I$;
	\item the norm of the truncated wave function $\Psi_I$ decreases rapidly (as $c^2_I$) when $I$ increases.
\end{itemize}
In addition, as a consequence of the first point, we note that the computation of $e_I$ becomes faster when $I$ increases.

%%%%%%%%%%%%%%%%%
\section{
\label{sec:MC}
Monte Carlo method}
%%%%%%%%%%%%%%%%%%%%%%%%%%%%%%%%%%

\subsection{Monte Carlo estimator}
%%%%%%%%%%%%%%%%%
\label{sec:3a}
To get an expression of $\Etwo$ suitable for MC simulations, 
the second-order contribution is recast as
\begin{equation}
	\Etwo = \sum_{I=1}^{\Ndet} p_I \qty( \frac{e_I }{p_I} ),
\label{e2mc}
\end{equation}
and is thus rewritten as the following MC estimator
\begin{equation}
	\Etwo = \expval{ \frac{e_I }{p_I} }_{p_{I}}.
\label{e2mc2}
\end{equation}
Here, $p_I$ is an arbitrary probability distribution. 
The optimal choice for $p_I$ is given by the zero-variance condition, i.e.
\begin{equation}
	p^\text{opt}_I =  \frac{e_I}{\Etwo}.
\end{equation}
Note that $e_I$ and $\Etwo$ being both negative, the probability distribution $p_I$ is positive, as it should be.

To build a reasonable approximation of $p_I$, we note that the magnitude of $e_I$, as expressed in Eq.~\eqref{e2_contrib}, is essentially given by the norm of the truncated wave function $\Psi_I$ (see Eq.~\eqref{trunc}). 
Thus, a natural choice for the probability distribution is
\begin{equation}
	p_I 
	= \frac{\braket{\Psi_I}}{\sum_{J=1}^{\Ndet} \braket{\Psi_J}}
	= \frac{\sum_{J=I}^{\Ndet} c^2_J }{\sum_{J=1}^{\Ndet} \sum_{K=J}^{\Ndet} c_K^2}.
\label{pI-sum}
\end{equation}
In our simulations, we have observed that summing totally or partially the squared coefficients in 
the numerator does not change significantly the statistical fluctuations.
As a consequence, we restrict the summation in Eq.~\eqref{pI-sum} to the leading term, i.e.
\begin{equation}
	p_I = \frac{c_I^2}{\sum_{J=1}^{\Ndet} c_J^2} = w_I.
\label{pI}
\end{equation}

\alert{
Let us emphasize that performing a MC simulation in the $e_I$ space 
is highly beneficial
since the number of $e_I$ is always small enough to make them all fit in memory.
Hence, one can follow the so-called \emph{lazy evaluation} strategy:\cite{Hudak}
the value of $e_I$ is computed only once when needed for the first time, and its value is then stored.  
If the same $e_I$ is requested later, the stored value will be returned.
}

%%%%%%%%%%%%%%%%%
\subsection{Improved Monte Carlo sampling}
\label{sec:3b}
%%%%%%%%%%%%%%%%%
The stochastic calculation of $\Etwo$, Eq.~\eqref{e2mc2}, can be done in a standard way
by sampling the probability distribution and averaging the successive values 
of $e_I/p_I$. In practice, the sampling can be realized by drawing, at each MC step, a uniform 
random number $u \in \qty[0,1]$ and selecting the 
determinant $\ket{I}$ verifying 
\begin{equation}
	R(I-1) \le u \le R(I),
\label{eq:invcdf}
\end{equation}
where $R$ is the cumulative distribution function of the probability 
distribution defined as
\begin{equation}
R(I)=\sum_{J=1}^I p_J,
\end{equation}
with $R(0)=0$. 

\alert{At this stage, it is useful to take advantage of the fact that, thanks to the way the $e_I$'s have been constructed, the quantity 
to be averaged, $e_{I}/p_{I}$, is a slowly-varying function 
of $I$ (providing that the small-scale fluctuations present at the level of individual determinants have been averaged out). 
This property, which is well illustrated by Fig.~\ref{fig2}, is shared by $p_I \sim c^2_I$, hence by the ratio $e_{I}/p_{I}$. 
Thus, an efficient way to reduce the statistical fluctuations consists in sampling {\it piece-wisely} $\mD$ by decomposing it into subdomains where 
the integrand is a slowly-varying function (see the justification of this statement after Eq.~\eqref{eq30}).}

To implement this idea, the interval $[0,1]$ 
is divided into $M$ equally-spaced intervals $\mU_k$ and a ``comb'' of correlated random numbers
\begin{equation}
u_k=\frac{k-1+u}{M} \qq{for} k=1,\ldots,M,
\label{uk}
\end{equation}
covering uniformly $[0,1]$ is created (where $u$ is a single uniform random number). 
At each MC step, a $M$-tuple of determinants $(I_1,I_2,\ldots,I_M)$ verifying
\begin{equation}
	R(I_k-1) \le u_k \le R(I_k) \qq{for} k=1,\ldots,M
\label{ri}
\end{equation}
is drawn. 

Defining $\mD_k$ as the subset of determinants $\ket{I_k}$ satisfying 
$R(I_k) \in \mU_k$, we introduce the following partition
\begin{equation}
\mD = \bigcup_{k=1}^{M} \mD_k \qq{with} \mD_k \cap \mD_l=\emptyset,\; \forall k \ne l,
\label{mD}
\end{equation}
and express $\Etwo$ as a sum of $M$ contributions
associated with each $\mD_k$:
\begin{equation}
\Etwo = \sum_{k=1}^M \sum_{I_k \in {\mD_k}} e_{I_k}.
\label{eqA}
\end{equation}
Using the process described above (Eqs.~\eqref{uk} and ~\eqref{ri}), the second-order 
energy
can be rewritten as the following MC estimator 
\begin{equation}
\label{E2-estimator}
\Etwo = \expval{ \frac{1}{M} \sum_{k=1}^M \frac{e_{I_k}}{p_{I_k}} }_{p(I_1,\ldots,I_M)},
\end{equation}
where $p(I_1,\ldots,I_M)$ denotes the normalized probability distribution corresponding to
Eqs.~\eqref{uk} and \eqref{ri}.
Equation~\eqref{E2-estimator} follows from the fact that, by construction, $p_{I_k}$ is the 
$k$-th marginal distribution of $p(I_1,\ldots,I_M)$
\begin{equation}
\sum_{I_1} \ldots \sum_{I_{k-1}} \sum_{I_{k+1}} \ldots \sum_{I_M} p(I_1,\ldots,I_M)= M  p_{I_k},
\label{eqB}
\end{equation}
with
\begin{equation}
	\sum_{I_k \in \mD_k} p_{I_k}=\frac{1}{M}.
	\label{eq30}
\end{equation}
\alert{By drawing determinants on separate subsets $\mD_k$, the sum
%, $M^{-1} \sum_{k=1}^M e_{I_k}/p_{I_k}$, 
to be averaged in Eq.~\eqref{E2-estimator}, 
is expected to fluctuate less than the very same sum computed by independently drawing determinants over $\mD$. 
This remarkable property can be explained as follows.
For large $M$, the fluctuations of the sum based on independent drawings behave as in any MC scheme, i.e.~as $M^{-1/2}$. 
Using a comb covering evenly (with weight $p_I$) the determinant space, the situation is different since the sum can now be seen as a Riemann sum over $\mD$ with a residual error behaving as $M^{-1}$.
As a consequence, the overall reduction in statistical noise resulting from the use of the comb is expected to be of the order of $\sqrt{M}$.
We emphasize that such an attractive feature is only observed because $e_{I}/p_{I}$ is a slowly-varying function of $I$ (as mentioned above).
In the opposite case, the gain would vanish. 
In the application on the \ce{F2} molecule presented below (see Fig.~\ref{fig5}), the numerical results confirm this: a decrease of about one order of magnitude in statistical error is obtained when using $M$=100.
Note that using a comb reduces the estimator's variance, but does not change the typical inverse square root behavior of the statistical error with respect to the
number of MC steps.}

Note that Eq.~\eqref{mD} is actually not correct when some determinants (first
and/or last determinant of a given subset) belong to more than one subset. Thus, special 
care has to be taken for determinants at the boundary of two subsets,
but this difficulty can be easily circumvented by formally duplicating each of
these determinants into copies with suitable weights.

%%%%%%%%%%%%%%
\subsection{Hybrid stochastic-deterministic scheme}
\label{sec:dynamical}
%%%%%%%%%%%%%%
In practice, because the first few determinants are responsible for the most significant contribution in Eq.~\eqref{e2mc}, it is advantageous not to sample the entire reference space but to remove from the stochastic sampling 
the leading determinants.  
Consequently, $\Etwo$ is split into a deterministic $\EtwoD$ and a stochastic  $\EtwoS$ component, such as
\begin{equation}
\begin{split}
	\Etwo 
	& = \EtwoD + \EtwoS
	\\
	& = \sum_{J \in \mDD} e_J +  \expval{ \frac{1}{M} \sum_{k=1}^M \frac{e_{I_k}}{p_{I_k}} }_{p(I_1,\ldots,I_M)},
\end{split}
\end{equation}
where $\mDD$ is the set of determinants in the deterministic space, and $\mD_\text{S} = \mD \setminus \mDD$ is its stochastic counterpart.

At a given point of the simulation, some determinants have been drawn, and some have not. 
If we keep track of the list of the drawn determinants, we can check periodically, 
for each $\mD_k$, whether or not all elements have been drawn at least once.
If that is the case, the full set of determinants is moved to $\mDD$ and the 
corresponding contribution  $\sum_{I_k} e_{I_k}$ is added to $\EtwoD$.
The statistical average and error bar are then updated accordingly. 
The expression of the $\Etwo$ estimator is now time-dependent and, at the $m$th MC step, the deterministic part is given by
\begin{equation}
	\EtwoD(m) = \sum_{k=1}^{M} \Theta_k(m) \sum_{I_k} e_{I_k}, 
\end{equation}
where 
\begin{equation}
	\Theta_k(m) = 
		\begin{cases}
			1,	&	\text{if $\mD_k \subset \mDD$ at step $m$},
			\\
			0,	&	\text{otherwise}.
		\end{cases}
\end{equation}
On the other hand, the stochastic part is now given by
\begin{equation}
	\EtwoS(m)= \frac{1}{M} \sum_{k=1}^M \qty[1-\Theta_k(m)] \sum_{I_k \in \mD_k} w_{I_k}^{(m)} \frac{e_{I_k}}{p_{I_k}},
\label{finalE2}
\end{equation}
where
\begin{equation}
	w_{I_k}^{(m)} = \frac{n_{I_k}^{(m)}}{\sum_{J_k \in \mD_k} n_{J_k}^{(m)}},
\end{equation}
and $n_{I_k}^{(m)}$ denotes the number of times the determinant $I_k$ has been drawn at 
iteration $m$. 

If desired, the calculation can be carried on until the stochastic part entirely vanishes.
In that case, all the determinants are in $\mDD$ and the exact 
value of $\Etwo$ is obtained with zero statistical fluctuations.

\alert{
Finally, to make sure that a given set $\mD_k$ does not stay in the stochastic part because a very small number of its determinants has not been drawn, we have implemented an additional step as follows.
At each MC iteration (where a new comb is created), the contribution $e_I$ of the first not-yet-sampled determinant (i.e.~corresponding to the smallest $I$ value in the sorted determinant list) is calculated and stored. 
By doing this, the convergence of the hybrid stochastic-deterministic estimator is significantly improved.
Moreover, after $\Ndet$ MC steps, it is now guaranteed that the exact deterministic value is reached.}

%\alert{In practice, we remarked that drawing all the determinants of the 
%largest $\mD_k$'s can take a very long time. To push
%the algorithm further, we introduced a deterministic filling of the $\mD_k$'s. 
%This simply consists in forcing the computation
%of the first non-computed determinant every time a new comb is sampled. 
%This guarantees that after $\Ndet$ samples, the
%exact value will be obtained and the convergence is enhanced, especially at the end of the run.
%In the numerical applications presented in the next section, a polynomial decay 
%of the statistical error is observed.
%}

%%%%%%%%%%%%%%
\subsection{Upper bound on the computational time}
%%%%%%%%%%%%%%

In the present method, the vast majority of the computational time is spent calculating $e_I$'s.
A crucial point which makes the algorithm particularly efficient is \alert{the lazy evaluation of these
quantities.}
This implies that, in practice, the stochastic calculation will never be longer than the time needed to compute all the individual $e_I$'s (i.e.~the time necessary to complete the fully-deterministic calculation) due to the negligible time required by the MC sampling \alert{(drawing 100 million random numbers takes less than 3 seconds on a single CPU core)}.

Finally, it is noteworthy that the final expression of $\Etwo$ can be very easily decomposed into (strictly) independent calculations. 
The algorithm presented here is thus embarrassingly parallel (see Sec.~\ref{sec:speedup} below).

%%% FIG 3 %%%
\begin{figure}
	\includegraphics[width=0.95\columnwidth]{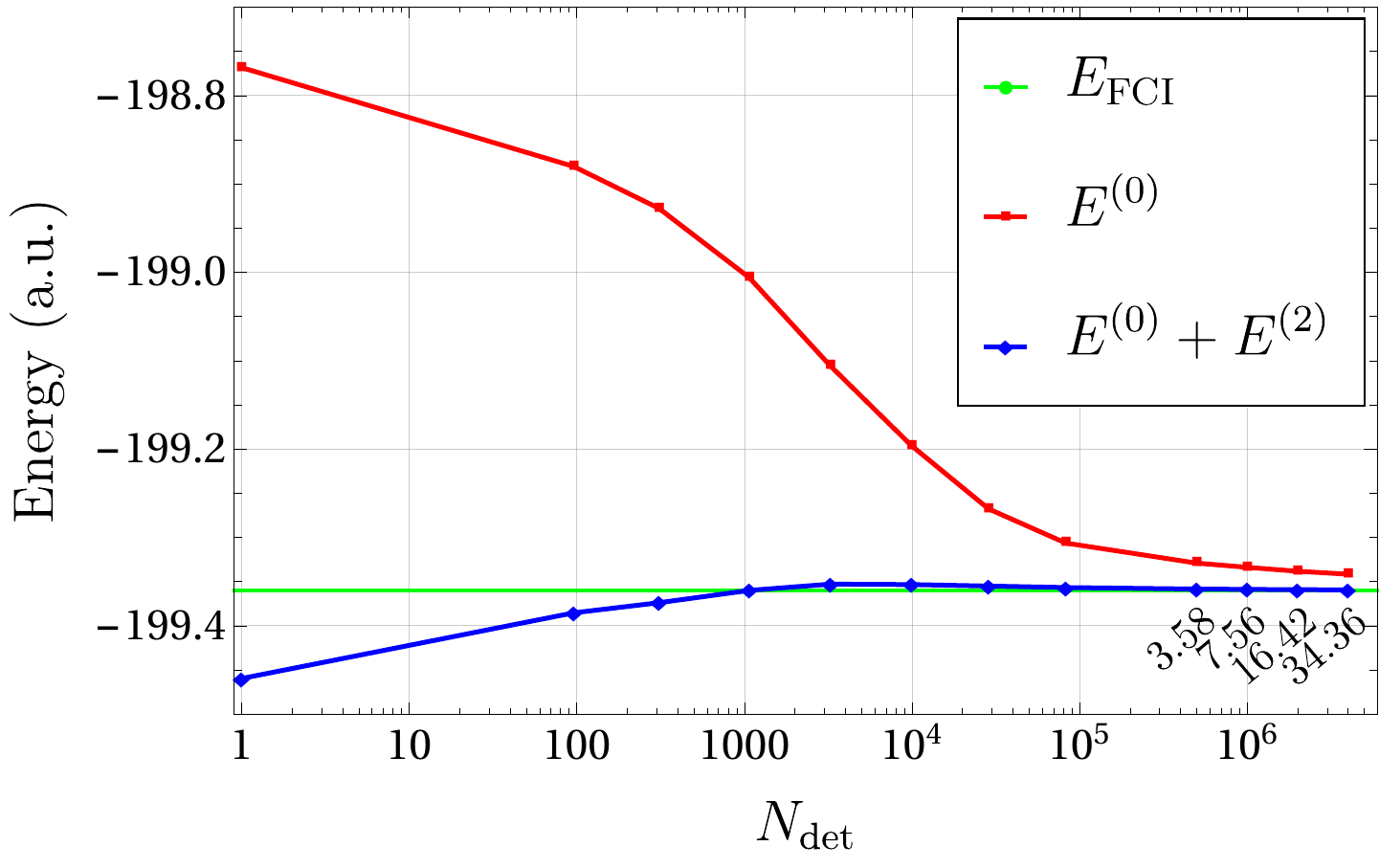}
\caption{
\alert{
\ce{F2} molecule at equilibrium geometry.
Convergence of the variational energy $\Ezero$ (red curve) as a function of the number of selected 
determinants $\Ndet$ obtained with the CIPSI method and the cc-pVQZ basis set.
The blue curve is obtained by adding the second-order energy contribution $\Etwo$
to the variational one $\Ezero$. 
The FCI value (green curve) is reported as a reference.
The wall-clock time (in minutes) needed to compute $\Etwo$ for 
various values of $\Ndet$ is also reported (black numbers underneath the blue curve).
}}
\label{fig3}
\end{figure}
%%%	%%%

%%% FIG 4 %%%
\begin{figure*}
	\includegraphics[height=0.3\linewidth]{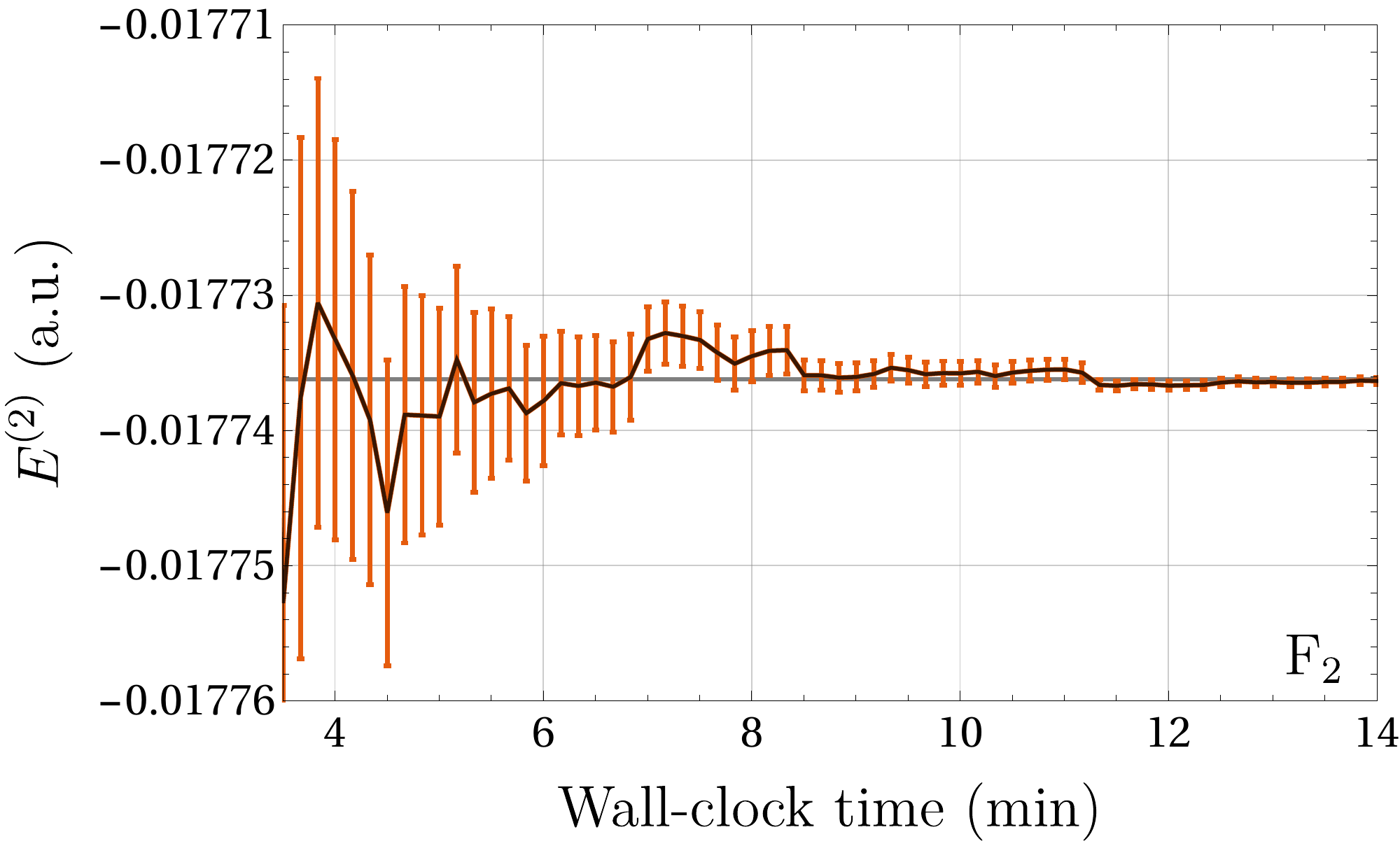}
	\includegraphics[height=0.3\linewidth]{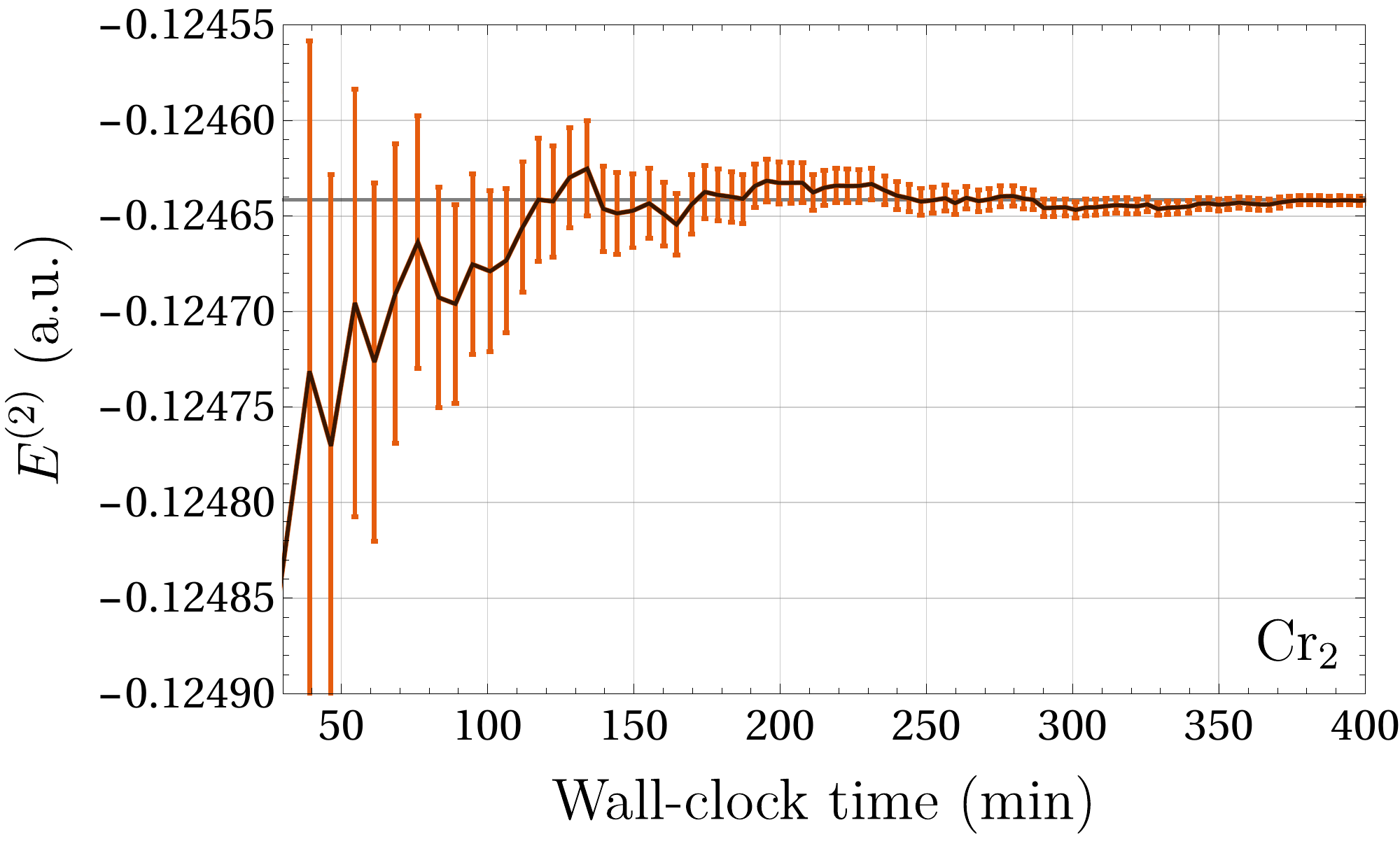}
\caption{Convergence of $\Etwo$ as a function of the wall-clock time for the \ce{F2} molecule (left) with $\Ndet = 4\times 10^6$ (cc-pVQZ basis set) and the \ce{Cr2} molecule (right) with {$\Ndet = 2 \times 10^7$ (cc-pVTZ basis set)}.
Both graphs are obtained with {800 CPU cores.}
The grey line corresponds to the exact (deterministic) value \alert{for \ce{F2} and to the value with the
lowest statistical error for \ce{Cr2}. 
The error bars correspond to one standard deviation.}}
\label{fig4}
\end{figure*}
%%%	%%%

%%% FIG 5 %%%
\begin{figure*}
	\includegraphics[width=.8\textwidth]{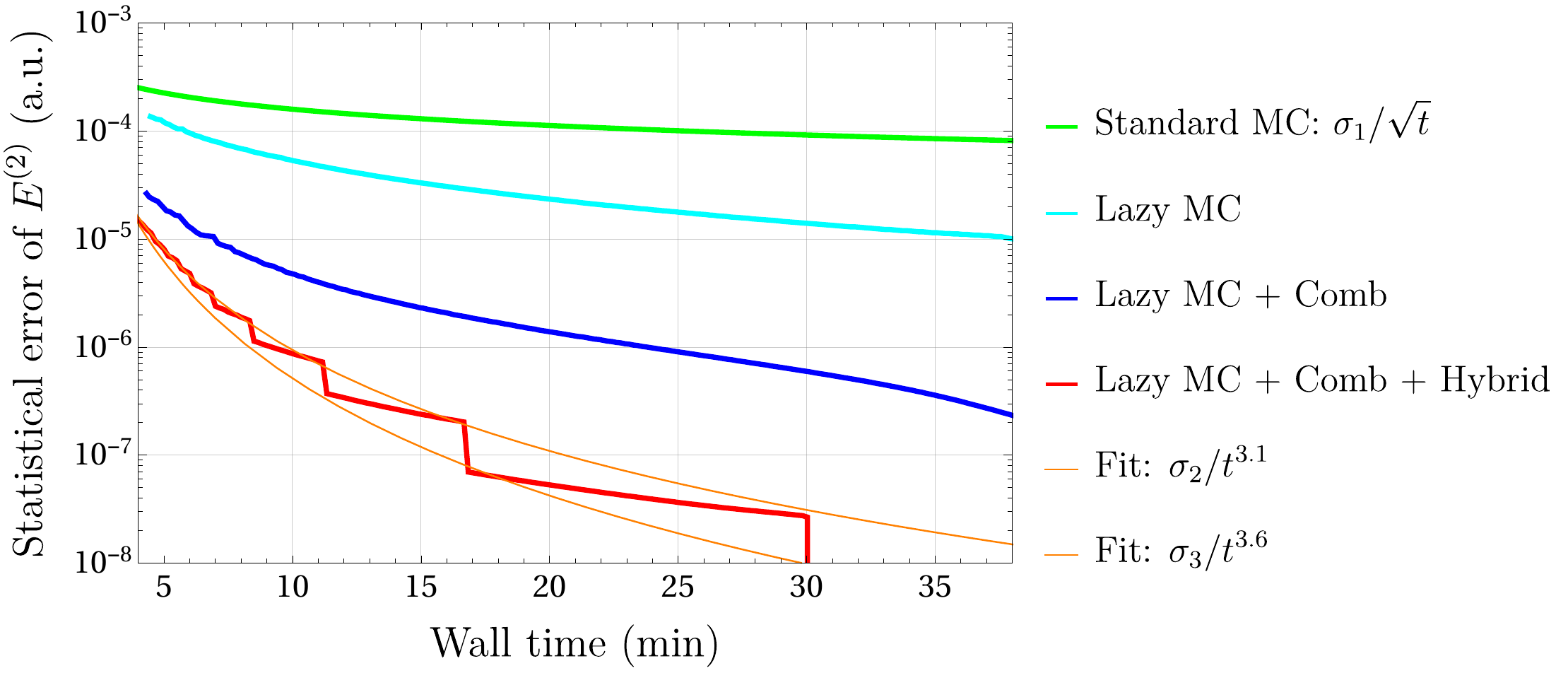}
\caption{\alert{Statistical error of $\Etwo$ as a function of the wall-clock time for the \ce{F2} molecule obtained with the cc-pVQZ basis and $\Ndet= 4\times 10^6$ with different schemes.}}
\label{fig5}
\end{figure*}
%%%	%%%

%%%%%%%%%%%%%%%%%
\section{
\label{sec:numerical}
Numerical tests}
%%%%%%%%%%%%%%%%%
\alert{
The present algorithm has been implemented in our Quantum Package code.\cite{quantum_package}
The perturbatively-selected CI algorithm CIPSI,\cite{huron_jcp_1973,cipsi_1983} as described in Ref.~\onlinecite{paci}, is used to build the multideterminant reference space.
In all the calculations performed in this section, we have chosen to use a comb with $M=100$. 
All the simulations were performed on the Curie supercomputer (TGCC/CEA/GENCI) where each node is a dual socket
Xeon E5-2680 @ 2.70GHz with 64GiB of RAM, interconnected with an Infiniband QDR network.
}

\subsection{F$_2 $ molecule}
%%%%%%%%%%%%%%%%%
As a first illustrative example, we consider the calculation of $\Etwo$ for the \ce{F2} molecule in its $^1\Sigma_g^+$ electronic ground state at equilibrium geometry. 
The two $1s$ core electrons are kept frozen and the Dunning's cc-pVQZ basis set is used.
The Hilbert space is built by distributing the 14 active electrons within the 108 non-frozen molecular orbitals for a total of more than $10^{20}$ determinants.

Despite the huge size of the Hilbert space, the selected CI approach is able to reach the {full CI (FCI)} limit with a very good accuracy. 
The convergence of the variational energy $\Ezero$ and of the total energy (given by the sum of the variational and second-order contribution $\Ezero+\Etwo$) with respect to the number of selected determinants are presented in Fig.~\ref{fig3}.
The maximum number of determinants we have selected is $4 \times 10^6$. 
For this value, $\Ezero$ is not converged but is already a 
reasonable approximation to the FCI energy with an error of about 18 \mEh. 
In sharp contrast, the total energy including the second-order correction 
converges very rapidly: millihartree accuracy 
is reached with about $2 \times 10^6$ determinants. 
For $\Ndet = 4 \times 10^6$, the best value obtained is $-199.3594$~{a.u.}, 
in quantitative agreement with the estimated FCI value of $-199.3598(2)$~{a.u.}~obtained by Cleland et al.~with FCIQMC\@.\cite{fciqmc2}

For this system and the maximum number of selected determinants considered, 
it is actually possible to calculate exactly $\Etwo$ by explicit evaluation of the entire sum 
(deterministic method).
The corresponding wall-clock times (in minutes) using 50 nodes (800 cores) are reported directly in Fig.~\ref{fig3}.
For $\Ndet= 10^4$, the calculation \alert{takes a few seconds while for the largest number of $\Ndet= 4 \times 10^6$ about 35 minutes are needed}.

We now consider the hybrid stochastic-deterministic evaluation of $\Etwo$. 
The left graph of Fig.~\ref{fig4} shows the evolution of $\Etwo$ as a function of the wall-clock time (in minutes). 
Data are given for the cc-pVQZ basis and $\Ndet= 4 \times 10^6$. 
Similar curves are obtained with the two other basis sets.
As one can see, the rate of convergence of the error is striking and, eventually, the exact value is obtained with very small fluctuations. 
If chemical accuracy is targeted (error of roughly 1 \mEh), three minutes are
needed using 800 cores. \alert{This value 
has to be compared with the $\sim 35$ minutes needed to evaluate
the exact value (see Fig.~\ref{fig3}).}

\alert{To have a better look at the fluctuations, the statistical error as a 
function of the wall-clock time is reported in Fig.~\ref{fig5}.
We have reported four curves to show the effects of the different strategies used in our algorithm.
The first one (in green) is the curve one would typically obtain using a standard MC algorithm where the contributions are always recomputed (no lazy evaluation). 
Note that, for this particular curve, we have not performed the calculation, but we have plotted an arbitrary $\sigma_1/\sqrt{t}$ curve to illustrate its decay rate.
The light blue curve is obtained using the MC estimator proposed in Sec.~\ref{sec:3a}. 
The slope is steeper than for the standard MC scheme thanks to the lazy evaluation strategy.
The introduction of the comb (Sec.~\ref{sec:3b}) reduces the statistical error by an order of magnitude, and produces the dark blue curve. 
Finally, incorporating the hybrid deterministic/stochastic scheme (Sec.~\ref{sec:dynamical}) yields the red curve.}
 
Quite remarkably, the overall convergence of the red curve is extremely rapid. 
Because of the irregular convergence, it is not easy to extract the exact mathematical form of the decay. 
However, it is clear that a typical \alert{polynomial} decay is observed.
\alert{Fitting the curve of the hybrid scheme gives a decrease of the error bar between $t^{-3.1}$ and $t^{-3.6}$, which is {significantly}
faster than the $t^{-1/2}$ behavior of the standard MC algorithm.}
Note also that some discontinuities in the statistical error are regularly observed. 
Such sudden drops occur each time a subset $\mD_k$ is entirely filled and its contribution is transferred to the deterministic part.
Comparison with standard MC illustrates that obtaining an arbitrary accuracy with a standard MC sampling can rapidly become prohibitively expensive. 
Most importantly, \alert{the wall-clock time} would rapidly become larger than the price to pay to compute exactly (i.e.~deterministically) $\Etwo$, \alert{which is not the case with the here-proposed method}.

%%% TABLE I %%%
\begin{table}
\caption{Convergence of $\Etwo$ for the \ce{Cr2} molecule with bond length $1.68$~\AA\, as a function of the wall-clock time for various basis sets {(800 CPU cores).}}

\begin{ruledtabular}
\begin{tabular}{llc}
Basis &  \mc{$\Etwo$} & Wall-clock time \\
\hline
cc-pVDZ &           $-0.068\,3(1)$  & 14 min     \\
        &           $-0.068\,36(1)$  & 55 min     \\
        &           $-0.068\,361(1)$  & 2.4 hr \\
        &           $-0.068\,360\,604$ & 3 hr \\
\hline
cc-pVTZ & $-0.124\,4(5)$  & 19 min \\
        & $-0.124\,7(1)$  & 58 min \\
        & $-0.124\,63(1)$  & 3.5 hr \\
        & $-0.124\,642(1)$   & 8.7 hr \\
        &  \qquad---   & $\sim$ 15 hr (estimated)\\
\hline
cc-pVQZ & $-0.155\,8(5)$    & 56 min \\
        & $-0.155\,9(1)$    & 2.5 hr \\
        & $-0.155\,95(1)$   & 9.0 hr \\
        & $-0.155\,952(1)$  & 18.5 hr \\
        &  \qquad--- &  $\sim$ 29 hr (estimated)\\
\end{tabular}
\end{ruledtabular}
\label{tab1}
\end{table}
%%%	%%%

%%% TABLE II %%%
\begin{table*}
\caption{Variational ground-state energy $\Ezero$ and second-order contribution $\Etwo$ of the \ce{Cr2} molecule with bond length $1.68$~\AA~computed with various basis sets. 
\alert{For all basis sets, the reference is composed of $2 \times 10^7$ determinants selected in the valence FCI space (28 electrons).}}
\begin{ruledtabular}
\begin{tabular}{lccrlll}
Reference  	&     Basis		&	Active space	&    \mc{$\Ezero$}    &   \mc{$\Etwo$}   & \mc{$\Ezero + \Etwo$}  \\ 
\hline
%CASSCF		&	cc-pVDZ		&  (12e,12o) 		
%&  $-2086.650\,896\,3$ & $-0.636\,57(5)$  &  $-2087.287\,47(5)$   \\ 
%			&	cc-pVTZ		&  (12e,12o) 		
%			&  $-2086.655\,594\,6$ & $-0.933\,05(5)$  
%   &  $-2087.588\,64(5)$   \\                  
%			&	cc-pVQZ		&  (12e,12o) 		&  $-2086.644\,989\,2$ & $-1.047\,34(4)$  &  $-2087.690\,55(4)$  \\
%\hline
CIPSI    		&	cc-pVDZ		&  (28e,76o) 		& $-2087.227\,883\,3$ & $-0.068\,334(1)$ &  $-2087.296\,217(1)$   \\ 
 			&	cc-pVTZ		&  (28e,126o) 		& $-2087.449\,781\,7$ & $-0.124\,676(1)$ &  $-2087.574\,423(1)$   \\ 
		  	&	cc-pVQZ		&  (28e,176o) 		& $-2087.513\,373\,3$ & $-0.155\,957(1)$ &  $-2087.669\,330(1)$   \\ 
\end{tabular}
\end{ruledtabular}
\label{tab2}
\end{table*}
%%%	%%%

%%%%%%%%%%%%%%%%%
\subsection{\ce{Cr2} molecule}
%%%%%%%%%%%%%%%%%
We now consider the challenging example of the \ce{Cr2} molecule in its $^1\Sigma_{\text{g}}^+$ ground state.
The internuclear distance is chosen to be close to its experimental equilibrium geometry, i.e.~$R_\text{Cr-Cr}$ = 1.68~\AA\@. 
Full-valence calculations including 28 active electrons (two frozen neon cores) are performed. 
The cc-pVDZ, TZ and QZ basis sets\cite{Balabanov_2005} are employed and the associated active spaces corresponding to (28e,76o), (28e,126o) and \alert{(28e,176o) include more than $10^{29}$, $10^{36}$ and $10^{42}$ determinants, respectively.
For all the basis sets, the \alert{molecular orbitals (MOs)} were obtained with the GAMESS \cite{gamess} program using a CASSCF calculation with 12 electrons in 12 orbitals, and $2 \times 10^7$ determinants were selected in the FCI space with the
CIPSI algorithm implemented in Quantum Package.
In the cc-pVQZ basis set, we had to remove the $h$ functions of the basis set since the version of GAMESS we used (prior to 2013) does not handle the corresponding two-electron integrals.}

\alert{The right graph of Fig.~\ref{fig4} shows the convergence of $\Etwo$ as a function of the wall-clock time for the cc-pVTZ basis set and $\Ndet = 2 \times 10^7$.}
Again, similar curves are obtained with the two other basis sets.
Similarly to \ce{F2}, the convergence is remarkably fast with a steep decrease of the statistical error with respect to the wall-clock time (for quantitative results, see Table~\ref{tab1}). 
Note that the maximum energy range in the right graph of Fig.~\ref{fig4} is only {0.35 \mEh}. 

Table~\ref{tab2} reports the quantitative results obtained with the {three} basis sets. 
One can see that very accurate results for $\Etwo$ can be obtained even with the largest QZ basis set. 
For the three basis sets, the statistical error obtained {is $10^{-6}$ \Eh.}
However, it is clear that in practical applications we do not need such high level of accuracy as the finite-size basis effects as well as the high-order perturbative contributions are much larger.
If, more reasonably, we target an accuracy of about 0.1 \mEh, we see in Table~\ref{tab1} that the wall-clock time needed is about \alert{14 min, 1 hr and 2.5 hr with 800 CPU cores} for the DZ, TZ, and QZ basis sets, respectively.
Finally, we note that, in contrast with \ce{F2}, the absolute value of $\Etwo$ remains large even when relatively large MR wave functions are employed.
This result clearly reflects the difficulty in treating accurately \ce{Cr2}. 
We postpone to a forthcoming paper the detailed analysis of this system and the calculation of the entire potential energy curve.

%%%%%%%%%%%%%%%%%
\subsection{Parallel speedup}
\label{sec:speedup}
%%%%%%%%%%%%%%%%%

%%% FIGURE VI %%%
\begin{figure}
	\includegraphics[width=0.9\columnwidth]{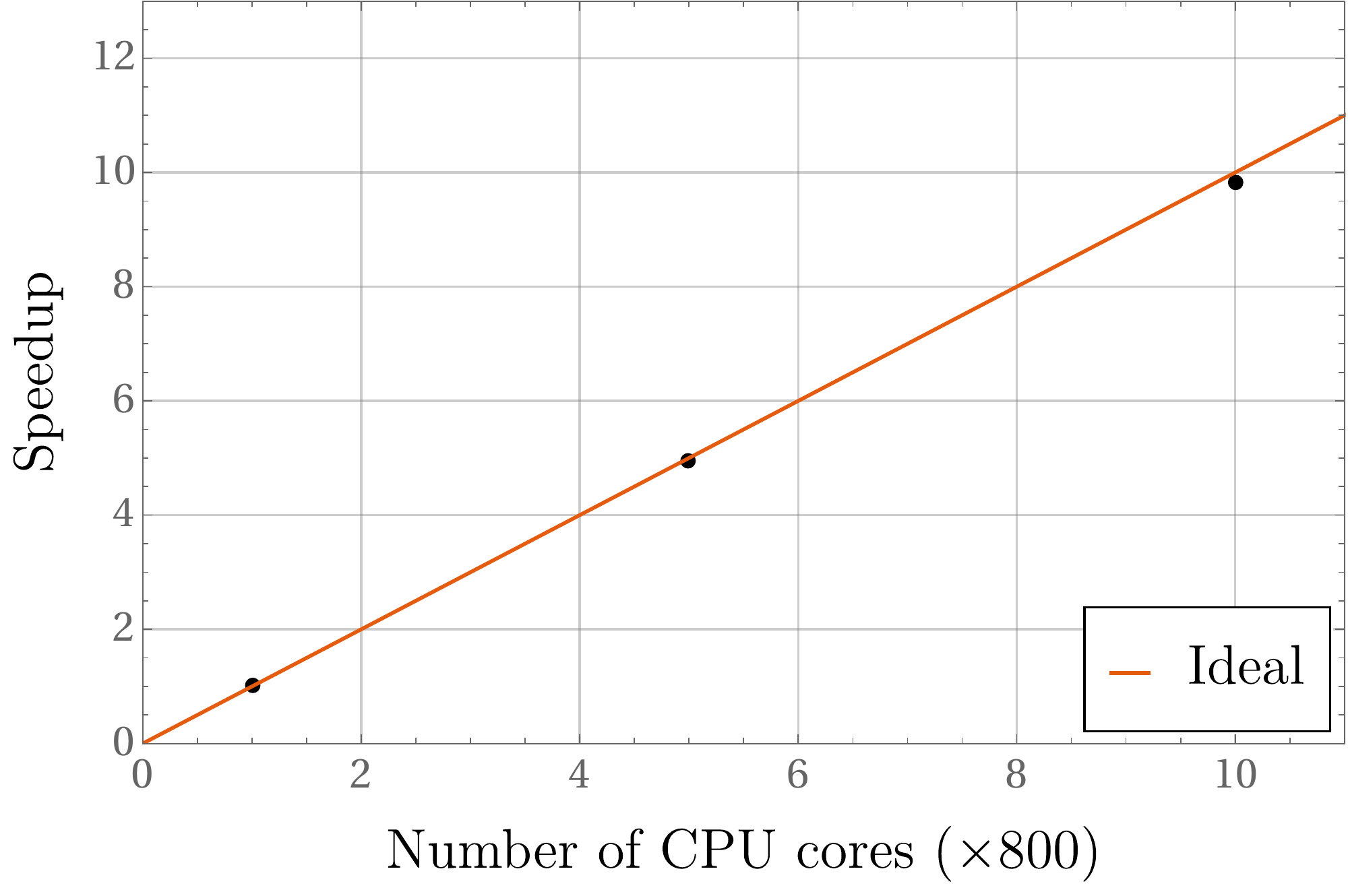}
\caption{Parallel speedup of our implementation using $800$, $4\,000$ and $8\,000$ cores.
The reference is the 800-core run.}
\label{fig6}
\end{figure}
%%%	%%%

\alert{To measure the parallel speedup of the present implementation of our algorithm, we have measured the wall-clock time needed to reach a
target statistical error of $10^{-6}$~a.u.~with $800$, $4\,000$ and $8\,000$ cores
($50$, $250$ and $500$ nodes) using the \ce{Cr2}/cc-pVQZ wave function with $\Ndet = 2 \times 10^7$. 
The speedup is calculated using the 800-{core} run as the reference, and the results
are shown in Fig.~\ref{fig6}. Going from $800$ to $4\,000$ cores gives a speedup of $4.95$, and
the $8\,000$-{core} run exhibits a speedup of $9.82$. 
These values are extremely close to the ideal
values of $5$ and $10$. 
Therefore, we believe that this method is a good candidate for running on exascale machines in a near future.}

%%%%%%%%%%%%%%%%%
\section{
\label{sec:conclusion}
Conclusions}
%%%%%%%%%%%%%%%%%
In this work, a hybrid stochastic-deterministic algorithm to compute 
the second-order energy contribution $\Etwo$ within Epstein-Nesbet MRPT has been introduced. 
Two main ideas are at the heart of the method. First, the reformulation of the 
standard expression of $\Etwo$, Eq.~\eqref{eq:pt2}, into Eq.~\eqref{pt2_final}. 
Thanks to the unique property of the elementary contributions $e_I$, the latter expression turns out to be particularly well-suited for low-variance MC calculations.
The second idea, which greatly enhances the convergence of the calculation, is to decompose $\Etwo$ as a sum of a deterministic and a stochastic part, the deterministic part being dynamically updated during the calculation.

We have observed that the size of the stochastic part (as well as the statistical error) decays in time with a \alert{polynomial} behavior.
If desired, the calculation can be carried on until the stochastic part entirely vanishes. 
In that case, the exact (deterministic) result is obtained with no error bar and no noticeable computational overhead compared to the fully-deterministic calculation.
Such a remarkable result is in sharp contrast with standard MC calculations where the statistical error decreases indefinitely as the inverse square root of the simulation time. 

The numerical applications presented for the \ce{F2} and \ce{Cr2} molecules illustrate the great efficiency of the method.
The largest calculation on \ce{Cr2} (cc-pVQZ basis set) has an active space of (28e,176o), corresponding to a Hilbert space consisting of approximately $10^{42}$ determinants and a multireference wave function containing $2 \times 10^7$ determinants. 
Even in this extreme case, $\Etwo$ can easily be calculated with sub-millihartree accuracy using a fully- and massively-parallel version of the algorithm.

%\alert{
%The current bottleneck of our implementation is memory. Each computing node
%has the full set of integrals in memory, as well as multiple copies of the wave
%function with different orderings. With $\Nbas = 186$ and $\Ndet=10^6$, the storage of a single instance of the wave function
%takes 5.3~GiB, so when multiple copies are stored the 64~GiB limit of a node is
%easily reached. We are currently working on the storage compression algorithm. 
%In the near future, taking advantage of massively parallel architectures, our aim is to perform routine calculations on wave functions with $\Ndet=10^9$.
%}

As a final comment, we would like to mention that, although we have only considered two 
illustrative examples in the present manuscript, our method has been shown to be 
highly successful in all the cases we have considered.

%%%%%%%%%%%%%%%%%
\begin{acknowledgments}
We thank the referees for their valuable comments on the first version of our manuscript.
This work was performed using HPC resources from CALMIP (Toulouse) under allocation 2016--0510 and from GENCI-TGCC (Grant 2016--08s015).
\end{acknowledgments}
%%%%%%%%%%%%%%%%%

%merlin.mbs aipnum4-1.bst 2010-07-25 4.21a (PWD, AO, DPC) hacked
%Control: key (0)
%Control: author (8) initials jnrlst
%Control: editor formatted (1) identically to author
%Control: production of article title (-1) disabled
%Control: page (0) single
%Control: year (1) truncated
%Control: production of eprint (0) enabled
%

\begin{thebibliography}{41}%
\makeatletter
\providecommand \@ifxundefined [1]{%
 \@ifx{#1\undefined}
}%
\providecommand \@ifnum [1]{%
 \ifnum #1\expandafter \@firstoftwo
 \else \expandafter \@secondoftwo
 \fi
}%
\providecommand \@ifx [1]{%
 \ifx #1\expandafter \@firstoftwo
 \else \expandafter \@secondoftwo
 \fi
}%
\providecommand \natexlab [1]{#1}%
\providecommand \enquote  [1]{``#1''}%
\providecommand \bibnamefont  [1]{#1}%
\providecommand \bibfnamefont [1]{#1}%
\providecommand \citenamefont [1]{#1}%
\providecommand \href@noop [0]{\@secondoftwo}%
\providecommand \href [0]{\begingroup \@sanitize@url \@href}%
\providecommand \@href[1]{\@@startlink{#1}\@@href}%
\providecommand \@@href[1]{\endgroup#1\@@endlink}%
\providecommand \@sanitize@url [0]{\catcode `\\12\catcode `\$12\catcode
  `\&12\catcode `\#12\catcode `\^12\catcode `\_12\catcode `\%12\relax}%
\providecommand \@@startlink[1]{}%
\providecommand \@@endlink[0]{}%
\providecommand \url  [0]{\begingroup\@sanitize@url \@url }%
\providecommand \@url [1]{\endgroup\@href {#1}{\urlprefix }}%
\providecommand \urlprefix  [0]{URL }%
\providecommand \Eprint [0]{\href }%
\providecommand \doibase [0]{http://dx.doi.org/}%
\providecommand \selectlanguage [0]{\@gobble}%
\providecommand \bibinfo  [0]{\@secondoftwo}%
\providecommand \bibfield  [0]{\@secondoftwo}%
\providecommand \translation [1]{[#1]}%
\providecommand \BibitemOpen [0]{}%
\providecommand \bibitemStop [0]{}%
\providecommand \bibitemNoStop [0]{.\EOS\space}%
\providecommand \EOS [0]{\spacefactor3000\relax}%
\providecommand \BibitemShut  [1]{\csname bibitem#1\endcsname}%
\let\auto@bib@innerbib\@empty
%</preamble>
\bibitem [{\citenamefont {Szalay}\ \emph {et~al.}(2012)\citenamefont {Szalay},
  \citenamefont {M{\"u}ller}, \citenamefont {Gidofalvi}, \citenamefont
  {Lischka},\ and\ \citenamefont {Shepard}}]{Szalay_2012}%
  \BibitemOpen
  \bibfield  {author} {\bibinfo {author} {\bibfnamefont {P.~G.}\ \bibnamefont
  {Szalay}}, \bibinfo {author} {\bibfnamefont {T.}~\bibnamefont {M{\"u}ller}},
  \bibinfo {author} {\bibfnamefont {G.}~\bibnamefont {Gidofalvi}}, \bibinfo
  {author} {\bibfnamefont {H.}~\bibnamefont {Lischka}}, \ and\ \bibinfo
  {author} {\bibfnamefont {R.}~\bibnamefont {Shepard}},\ }\href {\doibase
  10.1021/cr200137a} {\bibfield  {journal} {\bibinfo  {journal} {Chem. Rev.}\
  }\textbf {\bibinfo {volume} {112}},\ \bibinfo {pages} {108} (\bibinfo {year}
  {2012})}\BibitemShut {NoStop}%
\bibitem [{\citenamefont {H{\"a}ttig}\ \emph {et~al.}(2012)\citenamefont
  {H{\"a}ttig}, \citenamefont {Klopper}, \citenamefont {K{\"o}hn},\ and\
  \citenamefont {Tew}}]{Hattig_2012}%
  \BibitemOpen
  \bibfield  {author} {\bibinfo {author} {\bibfnamefont {C.}~\bibnamefont
  {H{\"a}ttig}}, \bibinfo {author} {\bibfnamefont {W.}~\bibnamefont {Klopper}},
  \bibinfo {author} {\bibfnamefont {A.}~\bibnamefont {K{\"o}hn}}, \ and\
  \bibinfo {author} {\bibfnamefont {D.~P.}\ \bibnamefont {Tew}},\ }\href
  {\doibase 10.1021/cr200168z} {\bibfield  {journal} {\bibinfo  {journal}
  {Chem. Rev.}\ }\textbf {\bibinfo {volume} {112}},\ \bibinfo {pages} {4}
  (\bibinfo {year} {2012})}\BibitemShut {NoStop}%
\bibitem [{\citenamefont {Lischka}\ \emph {et~al.}(1981)\citenamefont
  {Lischka}, \citenamefont {Shepard}, \citenamefont {Brown},\ and\
  \citenamefont {Shavitt}}]{mrci1}%
  \BibitemOpen
  \bibfield  {author} {\bibinfo {author} {\bibfnamefont {H.}~\bibnamefont
  {Lischka}}, \bibinfo {author} {\bibfnamefont {R.}~\bibnamefont {Shepard}},
  \bibinfo {author} {\bibfnamefont {F.~B.}\ \bibnamefont {Brown}}, \ and\
  \bibinfo {author} {\bibfnamefont {I.}~\bibnamefont {Shavitt}},\ }\href@noop
  {} {\bibfield  {journal} {\bibinfo  {journal} {Int. J. Quantum Chem.}\
  }\textbf {\bibinfo {volume} {20}},\ \bibinfo {pages} {91} (\bibinfo {year}
  {1981})}\BibitemShut {NoStop}%
\bibitem [{\citenamefont {Szalay}\ \emph {et~al.}(2011)\citenamefont {Szalay},
  \citenamefont {M{\"u}ller}, \citenamefont {Gidofalvi}, \citenamefont
  {Lischka},\ and\ \citenamefont {Shepard}}]{mrci2}%
  \BibitemOpen
  \bibfield  {author} {\bibinfo {author} {\bibfnamefont {P.~G.}\ \bibnamefont
  {Szalay}}, \bibinfo {author} {\bibfnamefont {T.}~\bibnamefont {M{\"u}ller}},
  \bibinfo {author} {\bibfnamefont {G.}~\bibnamefont {Gidofalvi}}, \bibinfo
  {author} {\bibfnamefont {H.}~\bibnamefont {Lischka}}, \ and\ \bibinfo
  {author} {\bibfnamefont {R.}~\bibnamefont {Shepard}},\ }\href@noop {}
  {\bibfield  {journal} {\bibinfo  {journal} {Chem. Rev.}\ }\textbf {\bibinfo
  {volume} {112}},\ \bibinfo {pages} {108} (\bibinfo {year}
  {2011})}\BibitemShut {NoStop}%
\bibitem [{\citenamefont {Lischka}\ \emph {et~al.}(2001)\citenamefont
  {Lischka}, \citenamefont {Shepard}, \citenamefont {Pitzer}, \citenamefont
  {Shavitt}, \citenamefont {Dallos}, \citenamefont {M{\"u}ller}, \citenamefont
  {Szalay}, \citenamefont {Seth}, \citenamefont {Kedziora}, \citenamefont
  {Yabushita},\ and\ \citenamefont {Zhang}}]{mrci3}%
  \BibitemOpen
  \bibfield  {author} {\bibinfo {author} {\bibfnamefont {H.}~\bibnamefont
  {Lischka}}, \bibinfo {author} {\bibfnamefont {R.}~\bibnamefont {Shepard}},
  \bibinfo {author} {\bibfnamefont {R.~M.}\ \bibnamefont {Pitzer}}, \bibinfo
  {author} {\bibfnamefont {I.}~\bibnamefont {Shavitt}}, \bibinfo {author}
  {\bibfnamefont {M.}~\bibnamefont {Dallos}}, \bibinfo {author} {\bibfnamefont
  {T.}~\bibnamefont {M{\"u}ller}}, \bibinfo {author} {\bibfnamefont {P.~G.}\
  \bibnamefont {Szalay}}, \bibinfo {author} {\bibfnamefont {M.}~\bibnamefont
  {Seth}}, \bibinfo {author} {\bibfnamefont {G.~S.}\ \bibnamefont {Kedziora}},
  \bibinfo {author} {\bibfnamefont {S.}~\bibnamefont {Yabushita}}, \ and\
  \bibinfo {author} {\bibfnamefont {Z.}~\bibnamefont {Zhang}},\ }\href@noop {}
  {\bibfield  {journal} {\bibinfo  {journal} {Phys. Chem. Chem. Phys.}\
  }\textbf {\bibinfo {volume} {3}},\ \bibinfo {pages} {664} (\bibinfo {year}
  {2001})}\BibitemShut {NoStop}%
\bibitem [{\citenamefont {Paldus}\ and\ \citenamefont {Li}(1999)}]{mrcc1}%
  \BibitemOpen
  \bibfield  {author} {\bibinfo {author} {\bibfnamefont {J.}~\bibnamefont
  {Paldus}}\ and\ \bibinfo {author} {\bibfnamefont {X.}~\bibnamefont {Li}},\
  }\href@noop {} {\bibfield  {journal} {\bibinfo  {journal} {Adv. Chem. Phys.}\
  ,\ \bibinfo {pages} {1}} (\bibinfo {year} {1999})}\BibitemShut {NoStop}%
\bibitem [{\citenamefont {Kong}\ \emph {et~al.}(2009)\citenamefont {Kong},
  \citenamefont {Shamasundar}, \citenamefont {Demel},\ and\ \citenamefont
  {Nooijen}}]{mrcc2}%
  \BibitemOpen
  \bibfield  {author} {\bibinfo {author} {\bibfnamefont {L.}~\bibnamefont
  {Kong}}, \bibinfo {author} {\bibfnamefont {K.}~\bibnamefont {Shamasundar}},
  \bibinfo {author} {\bibfnamefont {O.}~\bibnamefont {Demel}}, \ and\ \bibinfo
  {author} {\bibfnamefont {M.}~\bibnamefont {Nooijen}},\ }\href@noop {}
  {\bibfield  {journal} {\bibinfo  {journal} {J. Chem. Phys.}\ }\textbf
  {\bibinfo {volume} {130}},\ \bibinfo {pages} {114101} (\bibinfo {year}
  {2009})}\BibitemShut {NoStop}%
\bibitem [{\citenamefont {Jeziorski}(2010)}]{mrcc3}%
  \BibitemOpen
  \bibfield  {author} {\bibinfo {author} {\bibfnamefont {B.}~\bibnamefont
  {Jeziorski}},\ }\href@noop {} {\bibfield  {journal} {\bibinfo  {journal}
  {Mol. Phys.}\ }\textbf {\bibinfo {volume} {108}},\ \bibinfo {pages} {3043}
  (\bibinfo {year} {2010})}\BibitemShut {NoStop}%
\bibitem [{\citenamefont {Lyakh}\ \emph {et~al.}(2012)\citenamefont {Lyakh},
  \citenamefont {Musia{\l}}, \citenamefont {Lotrich},\ and\ \citenamefont
  {Bartlett}}]{Lyakh_2012}%
  \BibitemOpen
  \bibfield  {author} {\bibinfo {author} {\bibfnamefont {D.~I.}\ \bibnamefont
  {Lyakh}}, \bibinfo {author} {\bibfnamefont {M.}~\bibnamefont {Musia{\l}}},
  \bibinfo {author} {\bibfnamefont {V.~F.}\ \bibnamefont {Lotrich}}, \ and\
  \bibinfo {author} {\bibfnamefont {R.~J.}\ \bibnamefont {Bartlett}},\ }\href
  {\doibase 10.1021/cr2001417} {\bibfield  {journal} {\bibinfo  {journal}
  {Chem. Rev.}\ }\textbf {\bibinfo {volume} {112}},\ \bibinfo {pages} {182}
  (\bibinfo {year} {2012})}\BibitemShut {NoStop}%
\bibitem [{\citenamefont {Epstein}(1926)}]{epstein}%
  \BibitemOpen
  \bibfield  {author} {\bibinfo {author} {\bibfnamefont {S.}~\bibnamefont
  {Epstein}},\ }\href@noop {} {\bibfield  {journal} {\bibinfo  {journal} {Phys.
  Rev.}\ }\textbf {\bibinfo {volume} {28}},\ \bibinfo {pages} {695} (\bibinfo
  {year} {1926})}\BibitemShut {NoStop}%
\bibitem [{\citenamefont {Nesbet}(1955)}]{nesbet}%
  \BibitemOpen
  \bibfield  {author} {\bibinfo {author} {\bibfnamefont {R.~K.}\ \bibnamefont
  {Nesbet}},\ }\href@noop {} {\bibfield  {journal} {\bibinfo  {journal} {Proc.
  R. Soc. London, Ser. A}\ }\textbf {\bibinfo {volume} {230}},\ \bibinfo
  {pages} {312} (\bibinfo {year} {1955})}\BibitemShut {NoStop}%
\bibitem [{\citenamefont {Angeli}\ \emph {et~al.}(2001)\citenamefont {Angeli},
  \citenamefont {Cimiraglia}, \citenamefont {Evangelisti}, \citenamefont
  {Leininger},\ and\ \citenamefont {Malrieu}}]{nevpt1}%
  \BibitemOpen
  \bibfield  {author} {\bibinfo {author} {\bibfnamefont {C.}~\bibnamefont
  {Angeli}}, \bibinfo {author} {\bibfnamefont {R.}~\bibnamefont {Cimiraglia}},
  \bibinfo {author} {\bibfnamefont {S.}~\bibnamefont {Evangelisti}}, \bibinfo
  {author} {\bibfnamefont {T.}~\bibnamefont {Leininger}}, \ and\ \bibinfo
  {author} {\bibfnamefont {J.-P.}\ \bibnamefont {Malrieu}},\ }\href@noop {}
  {\bibfield  {journal} {\bibinfo  {journal} {J. Chem. Phys.}\ }\textbf
  {\bibinfo {volume} {114}},\ \bibinfo {pages} {10252} (\bibinfo {year}
  {2001})}\BibitemShut {NoStop}%
\bibitem [{\citenamefont {Angeli}, \citenamefont {Cimiraglia},\ and\
  \citenamefont {Malrieu}(2002)}]{nevpt2}%
  \BibitemOpen
  \bibfield  {author} {\bibinfo {author} {\bibfnamefont {C.}~\bibnamefont
  {Angeli}}, \bibinfo {author} {\bibfnamefont {R.}~\bibnamefont {Cimiraglia}},
  \ and\ \bibinfo {author} {\bibfnamefont {J.-P.}\ \bibnamefont {Malrieu}},\
  }\href@noop {} {\bibfield  {journal} {\bibinfo  {journal} {J. Chem. Phys.}\
  }\textbf {\bibinfo {volume} {117}},\ \bibinfo {pages} {9138} (\bibinfo {year}
  {2002})}\BibitemShut {NoStop}%
\bibitem [{\citenamefont {Fink}(2006)}]{fink1}%
  \BibitemOpen
  \bibfield  {author} {\bibinfo {author} {\bibfnamefont {R.~F.}\ \bibnamefont
  {Fink}},\ }\href@noop {} {\bibfield  {journal} {\bibinfo  {journal} {Chem.
  Phys. Lett.}\ }\textbf {\bibinfo {volume} {428}},\ \bibinfo {pages} {461}
  (\bibinfo {year} {2006})}\BibitemShut {NoStop}%
\bibitem [{\citenamefont {Fink}(2009)}]{fink2}%
  \BibitemOpen
  \bibfield  {author} {\bibinfo {author} {\bibfnamefont {R.~F.}\ \bibnamefont
  {Fink}},\ }\href@noop {} {\bibfield  {journal} {\bibinfo  {journal} {Chem.
  Phys.}\ }\textbf {\bibinfo {volume} {356}},\ \bibinfo {pages} {39} (\bibinfo
  {year} {2009})}\BibitemShut {NoStop}%
\bibitem [{\citenamefont {Andersson}\ \emph {et~al.}(1990)\citenamefont
  {Andersson}, \citenamefont {Malmqvist}, \citenamefont {Roos}, \citenamefont
  {Sadlej},\ and\ \citenamefont {Wolinski}}]{caspt1}%
  \BibitemOpen
  \bibfield  {author} {\bibinfo {author} {\bibfnamefont {K.}~\bibnamefont
  {Andersson}}, \bibinfo {author} {\bibfnamefont {P.~A.}\ \bibnamefont
  {Malmqvist}}, \bibinfo {author} {\bibfnamefont {B.~O.}\ \bibnamefont {Roos}},
  \bibinfo {author} {\bibfnamefont {A.~J.}\ \bibnamefont {Sadlej}}, \ and\
  \bibinfo {author} {\bibfnamefont {K.}~\bibnamefont {Wolinski}},\ }\href@noop
  {} {\bibfield  {journal} {\bibinfo  {journal} {J. Phys. Chem.}\ }\textbf
  {\bibinfo {volume} {94}},\ \bibinfo {pages} {5483} (\bibinfo {year}
  {1990})}\BibitemShut {NoStop}%
\bibitem [{\citenamefont {Andersson}, \citenamefont {Malmqvist},\ and\
  \citenamefont {Roos}(1992)}]{caspt2}%
  \BibitemOpen
  \bibfield  {author} {\bibinfo {author} {\bibfnamefont {K.}~\bibnamefont
  {Andersson}}, \bibinfo {author} {\bibfnamefont {P.-A.}\ \bibnamefont
  {Malmqvist}}, \ and\ \bibinfo {author} {\bibfnamefont {B.~O.}\ \bibnamefont
  {Roos}},\ }\href@noop {} {\bibfield  {journal} {\bibinfo  {journal} {J. Chem.
  Phys.}\ }\textbf {\bibinfo {volume} {96}},\ \bibinfo {pages} {1218} (\bibinfo
  {year} {1992})}\BibitemShut {NoStop}%
\bibitem [{\citenamefont {Siegbahn}\ \emph {et~al.}(1980)\citenamefont
  {Siegbahn}, \citenamefont {Heiberg}, \citenamefont {Roos},\ and\
  \citenamefont {L{\'e}vy}}]{casscf1}%
  \BibitemOpen
  \bibfield  {author} {\bibinfo {author} {\bibfnamefont {P.}~\bibnamefont
  {Siegbahn}}, \bibinfo {author} {\bibfnamefont {A.}~\bibnamefont {Heiberg}},
  \bibinfo {author} {\bibfnamefont {B.}~\bibnamefont {Roos}}, \ and\ \bibinfo
  {author} {\bibfnamefont {B.}~\bibnamefont {L{\'e}vy}},\ }\href@noop {}
  {\bibfield  {journal} {\bibinfo  {journal} {Physica Scripta}\ }\textbf
  {\bibinfo {volume} {21}},\ \bibinfo {pages} {323} (\bibinfo {year}
  {1980})}\BibitemShut {NoStop}%
\bibitem [{\citenamefont {Roos}, \citenamefont {Taylor},\ and\ \citenamefont
  {Siegbahn}(1980)}]{casscf2}%
  \BibitemOpen
  \bibfield  {author} {\bibinfo {author} {\bibfnamefont {B.~O.}\ \bibnamefont
  {Roos}}, \bibinfo {author} {\bibfnamefont {P.~R.}\ \bibnamefont {Taylor}}, \
  and\ \bibinfo {author} {\bibfnamefont {P.~E.~M.}\ \bibnamefont {Siegbahn}},\
  }\href@noop {} {\bibfield  {journal} {\bibinfo  {journal} {Chem. Phys.}\
  }\textbf {\bibinfo {volume} {48}},\ \bibinfo {pages} {157} (\bibinfo {year}
  {1980})}\BibitemShut {NoStop}%
\bibitem [{\citenamefont {Siegbahn}\ \emph {et~al.}(1981)\citenamefont
  {Siegbahn}, \citenamefont {Alml{\"o}f}, \citenamefont {Heiberg},\ and\
  \citenamefont {Roos}}]{casscf3}%
  \BibitemOpen
  \bibfield  {author} {\bibinfo {author} {\bibfnamefont {P.~E.~M.}\
  \bibnamefont {Siegbahn}}, \bibinfo {author} {\bibfnamefont {J.}~\bibnamefont
  {Alml{\"o}f}}, \bibinfo {author} {\bibfnamefont {A.}~\bibnamefont {Heiberg}},
  \ and\ \bibinfo {author} {\bibfnamefont {B.~O.}\ \bibnamefont {Roos}},\
  }\href@noop {} {\bibfield  {journal} {\bibinfo  {journal} {J. Chem. Phys.}\
  }\textbf {\bibinfo {volume} {74}},\ \bibinfo {pages} {2384} (\bibinfo {year}
  {1981})}\BibitemShut {NoStop}%
\bibitem [{\citenamefont {Huron}, \citenamefont {Rancurel},\ and\ \citenamefont
  {Malrieu}(1973)}]{huron_jcp_1973}%
  \BibitemOpen
  \bibfield  {author} {\bibinfo {author} {\bibfnamefont {B.}~\bibnamefont
  {Huron}}, \bibinfo {author} {\bibfnamefont {P.}~\bibnamefont {Rancurel}}, \
  and\ \bibinfo {author} {\bibfnamefont {J.~P.}\ \bibnamefont {Malrieu}},\
  }\href@noop {} {\bibfield  {journal} {\bibinfo  {journal} {J. Chem. Phys.}\
  }\textbf {\bibinfo {volume} {58}},\ \bibinfo {pages} {5745} (\bibinfo {year}
  {1973})}\BibitemShut {NoStop}%
\bibitem [{\citenamefont {Caffarel}\ \emph {et~al.}(2016)\citenamefont
  {Caffarel}, \citenamefont {Applencourt}, \citenamefont {Giner},\ and\
  \citenamefont {Scemama}}]{paci}%
  \BibitemOpen
  \bibfield  {author} {\bibinfo {author} {\bibfnamefont {M.}~\bibnamefont
  {Caffarel}}, \bibinfo {author} {\bibfnamefont {T.}~\bibnamefont
  {Applencourt}}, \bibinfo {author} {\bibfnamefont {E.}~\bibnamefont {Giner}},
  \ and\ \bibinfo {author} {\bibfnamefont {A.}~\bibnamefont {Scemama}},\
  }\enquote {\bibinfo {title} {Using cipsi nodes in diffusion monte carlo},}\
  in\ \href {\doibase 10.1021/bk-2016-1234.ch002} {\emph {\bibinfo {booktitle}
  {Recent Progress in Quantum Monte Carlo}}}\ (\bibinfo  {publisher} {American
  Chemical Society ({ACS})},\ \bibinfo {year} {2016})\ Chap.~\bibinfo {chapter}
  {2}, pp.\ \bibinfo {pages} {15--46},\ \Eprint
  {http://arxiv.org/abs/http://pubs.acs.org/doi/pdf/10.1021/bk-2016-1234.ch002}
  {http://pubs.acs.org/doi/pdf/10.1021/bk-2016-1234.ch002} \BibitemShut
  {NoStop}%
\bibitem [{\citenamefont {Booth}, \citenamefont {Thom},\ and\ \citenamefont
  {Alavi}(2009)}]{fciqmc1}%
  \BibitemOpen
  \bibfield  {author} {\bibinfo {author} {\bibfnamefont {G.~H.}\ \bibnamefont
  {Booth}}, \bibinfo {author} {\bibfnamefont {A.~J.~W.}\ \bibnamefont {Thom}},
  \ and\ \bibinfo {author} {\bibfnamefont {A.}~\bibnamefont {Alavi}},\
  }\href@noop {} {\bibfield  {journal} {\bibinfo  {journal} {J. Chem. Phys.}\
  }\textbf {\bibinfo {volume} {131}},\ \bibinfo {pages} {054106} (\bibinfo
  {year} {2009})}\BibitemShut {NoStop}%
\bibitem [{\citenamefont {Cleland}\ \emph {et~al.}(2012)\citenamefont
  {Cleland}, \citenamefont {Booth}, \citenamefont {Overy},\ and\ \citenamefont
  {Alavi}}]{fciqmc2}%
  \BibitemOpen
  \bibfield  {author} {\bibinfo {author} {\bibfnamefont {D.}~\bibnamefont
  {Cleland}}, \bibinfo {author} {\bibfnamefont {G.~H.}\ \bibnamefont {Booth}},
  \bibinfo {author} {\bibfnamefont {C.}~\bibnamefont {Overy}}, \ and\ \bibinfo
  {author} {\bibfnamefont {A.}~\bibnamefont {Alavi}},\ }\href@noop {}
  {\bibfield  {journal} {\bibinfo  {journal} {J. Chem. Theory Comput.}\
  }\textbf {\bibinfo {volume} {8}},\ \bibinfo {pages} {4138} (\bibinfo {year}
  {2012})}\BibitemShut {NoStop}%
\bibitem [{\citenamefont {Petruzielo}\ \emph {et~al.}(2012)\citenamefont
  {Petruzielo}, \citenamefont {Holmes}, \citenamefont {Changlani},
  \citenamefont {Nightingale},\ and\ \citenamefont {Umrigar}}]{fciqmc3}%
  \BibitemOpen
  \bibfield  {author} {\bibinfo {author} {\bibfnamefont {F.~R.}\ \bibnamefont
  {Petruzielo}}, \bibinfo {author} {\bibfnamefont {A.~A.}\ \bibnamefont
  {Holmes}}, \bibinfo {author} {\bibfnamefont {H.~J.}\ \bibnamefont
  {Changlani}}, \bibinfo {author} {\bibfnamefont {M.~P.}\ \bibnamefont
  {Nightingale}}, \ and\ \bibinfo {author} {\bibfnamefont {C.~J.}\ \bibnamefont
  {Umrigar}},\ }\href@noop {} {\bibfield  {journal} {\bibinfo  {journal} {Phys.
  Rev. Lett.}\ }\textbf {\bibinfo {volume} {109}},\ \bibinfo {pages} {230201}
  (\bibinfo {year} {2012})}\BibitemShut {NoStop}%
\bibitem [{\citenamefont {White}(1992)}]{dmrg1}%
  \BibitemOpen
  \bibfield  {author} {\bibinfo {author} {\bibfnamefont {S.~R.}\ \bibnamefont
  {White}},\ }\href@noop {} {\bibfield  {journal} {\bibinfo  {journal} {Phys.
  Rev. Lett.}\ }\textbf {\bibinfo {volume} {69}},\ \bibinfo {pages} {2863}
  (\bibinfo {year} {1992})}\BibitemShut {NoStop}%
\bibitem [{\citenamefont {White}(1993)}]{dmrg2}%
  \BibitemOpen
  \bibfield  {author} {\bibinfo {author} {\bibfnamefont {S.~R.}\ \bibnamefont
  {White}},\ }\href@noop {} {\bibfield  {journal} {\bibinfo  {journal} {Phys.
  Rev. B}\ }\textbf {\bibinfo {volume} {48}},\ \bibinfo {pages} {10345}
  (\bibinfo {year} {1993})}\BibitemShut {NoStop}%
\bibitem [{\citenamefont {Sharma}\ and\ \citenamefont {Chan}(2012)}]{dmrg3}%
  \BibitemOpen
  \bibfield  {author} {\bibinfo {author} {\bibfnamefont {S.}~\bibnamefont
  {Sharma}}\ and\ \bibinfo {author} {\bibfnamefont {G.-L.}\ \bibnamefont
  {Chan}},\ }\href@noop {} {\bibfield  {journal} {\bibinfo  {journal} {J. Chem.
  Phys.}\ }\textbf {\bibinfo {volume} {136}},\ \bibinfo {pages} {124121}
  (\bibinfo {year} {2012})}\BibitemShut {NoStop}%
\bibitem [{\citenamefont {{Giner}}\ \emph {et~al.}(2017)\citenamefont
  {{Giner}}, \citenamefont {{Angeli}}, \citenamefont {{Garniron}},
  \citenamefont {{Scemama}},\ and\ \citenamefont {{Malrieu}}}]{Giner_2017}%
  \BibitemOpen
  \bibfield  {author} {\bibinfo {author} {\bibfnamefont {E.}~\bibnamefont
  {{Giner}}}, \bibinfo {author} {\bibfnamefont {C.}~\bibnamefont {{Angeli}}},
  \bibinfo {author} {\bibfnamefont {Y.}~\bibnamefont {{Garniron}}}, \bibinfo
  {author} {\bibfnamefont {A.}~\bibnamefont {{Scemama}}}, \ and\ \bibinfo
  {author} {\bibfnamefont {J.-P.}\ \bibnamefont {{Malrieu}}},\ }\href@noop {}
  {\bibfield  {journal} {\bibinfo  {journal} {J. Chem. Phys.}\ }\textbf
  {\bibinfo {volume} {in press}} (\bibinfo {year} {2017})}\BibitemShut
  {NoStop}%
\bibitem [{\citenamefont {Evangelisti}, \citenamefont {Daudey},\ and\
  \citenamefont {Malrieu}(1983)}]{cipsi_1983}%
  \BibitemOpen
  \bibfield  {author} {\bibinfo {author} {\bibfnamefont {S.}~\bibnamefont
  {Evangelisti}}, \bibinfo {author} {\bibfnamefont {J.~P.}\ \bibnamefont
  {Daudey}}, \ and\ \bibinfo {author} {\bibfnamefont {J.~P.}\ \bibnamefont
  {Malrieu}},\ }\href@noop {} {\bibfield  {journal} {\bibinfo  {journal} {Chem.
  Phys.}\ }\textbf {\bibinfo {volume} {75}},\ \bibinfo {pages} {91} (\bibinfo
  {year} {1983})}\BibitemShut {NoStop}%
\bibitem [{\citenamefont {Willow}, \citenamefont {Kim},\ and\ \citenamefont
  {Hirata}(2012)}]{hirata1}%
  \BibitemOpen
  \bibfield  {author} {\bibinfo {author} {\bibfnamefont {S.~Y.}\ \bibnamefont
  {Willow}}, \bibinfo {author} {\bibfnamefont {K.~S.}\ \bibnamefont {Kim}}, \
  and\ \bibinfo {author} {\bibfnamefont {S.}~\bibnamefont {Hirata}},\
  }\href@noop {} {\bibfield  {journal} {\bibinfo  {journal} {J. Chem. Phys.}\
  }\textbf {\bibinfo {volume} {137}},\ \bibinfo {pages} {204122} (\bibinfo
  {year} {2012})}\BibitemShut {NoStop}%
\bibitem [{\citenamefont {Willow}\ \emph {et~al.}(2014)\citenamefont {Willow},
  \citenamefont {Zhang}, \citenamefont {Valeev},\ and\ \citenamefont
  {Hirata}}]{hirata2}%
  \BibitemOpen
  \bibfield  {author} {\bibinfo {author} {\bibfnamefont {S.~Y.}\ \bibnamefont
  {Willow}}, \bibinfo {author} {\bibfnamefont {J.}~\bibnamefont {Zhang}},
  \bibinfo {author} {\bibfnamefont {E.~F.}\ \bibnamefont {Valeev}}, \ and\
  \bibinfo {author} {\bibfnamefont {S.}~\bibnamefont {Hirata}},\ }\href@noop {}
  {\bibfield  {journal} {\bibinfo  {journal} {J. Chem. Phys.}\ }\textbf
  {\bibinfo {volume} {140}},\ \bibinfo {pages} {031101} (\bibinfo {year}
  {2014})}\BibitemShut {NoStop}%
\bibitem [{\citenamefont {Sharma}\ \emph {et~al.}(2017)\citenamefont {Sharma},
  \citenamefont {Holmes}, \citenamefont {Jeanmairet}, \citenamefont {Alavi},\
  and\ \citenamefont {Umrigar}}]{sharma1}%
  \BibitemOpen
  \bibfield  {author} {\bibinfo {author} {\bibfnamefont {S.}~\bibnamefont
  {Sharma}}, \bibinfo {author} {\bibfnamefont {A.~A.}\ \bibnamefont {Holmes}},
  \bibinfo {author} {\bibfnamefont {G.}~\bibnamefont {Jeanmairet}}, \bibinfo
  {author} {\bibfnamefont {A.}~\bibnamefont {Alavi}}, \ and\ \bibinfo {author}
  {\bibfnamefont {C.~J.}\ \bibnamefont {Umrigar}},\ }\href {\doibase
  10.1021/acs.jctc.6b01028} {\bibfield  {journal} {\bibinfo  {journal} {Journal
  of Chemical Theory and Computation}\ }\textbf {\bibinfo {volume} {13}},\
  \bibinfo {pages} {1595} (\bibinfo {year} {2017})},\ \bibinfo {note} {pMID:
  28263594},\ \Eprint
  {http://arxiv.org/abs/http://dx.doi.org/10.1021/acs.jctc.6b01028}
  {http://dx.doi.org/10.1021/acs.jctc.6b01028} \BibitemShut {NoStop}%
\bibitem [{\citenamefont {Jeanmairet}, \citenamefont {Sharma},\ and\
  \citenamefont {Alavi}(2017)}]{sharma2}%
  \BibitemOpen
  \bibfield  {author} {\bibinfo {author} {\bibfnamefont {G.}~\bibnamefont
  {Jeanmairet}}, \bibinfo {author} {\bibfnamefont {S.}~\bibnamefont {Sharma}},
  \ and\ \bibinfo {author} {\bibfnamefont {A.}~\bibnamefont {Alavi}},\ }\href
  {\doibase 10.1063/1.4974177} {\bibfield  {journal} {\bibinfo  {journal} {The
  Journal of Chemical Physics}\ }\textbf {\bibinfo {volume} {146}},\ \bibinfo
  {pages} {044107} (\bibinfo {year} {2017})},\ \Eprint
  {http://arxiv.org/abs/http://dx.doi.org/10.1063/1.4974177}
  {http://dx.doi.org/10.1063/1.4974177} \BibitemShut {NoStop}%
\bibitem [{\citenamefont {Szabo}\ and\ \citenamefont
  {Ostlund}(1989)}]{SzaboBook}%
  \BibitemOpen
  \bibfield  {author} {\bibinfo {author} {\bibfnamefont {A.}~\bibnamefont
  {Szabo}}\ and\ \bibinfo {author} {\bibfnamefont {N.~S.}\ \bibnamefont
  {Ostlund}},\ }\href@noop {} {\emph {\bibinfo {title} {Modern quantum
  chemistry}}}\ (\bibinfo  {publisher} {McGraw-Hill},\ \bibinfo {address} {New
  York},\ \bibinfo {year} {1989})\BibitemShut {NoStop}%
\bibitem [{\citenamefont {Scemama}\ and\ \citenamefont
  {Giner}(2013)}]{scemama_2013_3}%
  \BibitemOpen
  \bibfield  {author} {\bibinfo {author} {\bibfnamefont {A.}~\bibnamefont
  {Scemama}}\ and\ \bibinfo {author} {\bibfnamefont {E.}~\bibnamefont
  {Giner}},\ }\href {http://fr.arxiv.org/abs/1311.6244} {\bibfield  {journal}
  {\bibinfo  {journal} {arXiv:1311.6244 [physics.comp-ph]}\ } (\bibinfo {year}
  {2013})}\BibitemShut {NoStop}%
\bibitem [{\citenamefont {Dunning}(1989)}]{Dunning_1989}%
  \BibitemOpen
  \bibfield  {author} {\bibinfo {author} {\bibfnamefont {T.~H.}\ \bibnamefont
  {Dunning}},\ }\href {\doibase 10.1063/1.456153} {\bibfield  {journal}
  {\bibinfo  {journal} {J. Chem. Phys.}\ }\textbf {\bibinfo {volume} {90}},\
  \bibinfo {pages} {1007} (\bibinfo {year} {1989})}\BibitemShut {NoStop}%
\bibitem [{\citenamefont {Hudak}(1989)}]{Hudak}%
  \BibitemOpen
  \bibfield  {author} {\bibinfo {author} {\bibfnamefont {P.}~\bibnamefont
  {Hudak}},\ }\href@noop {} {\bibfield  {journal} {\bibinfo  {journal} {ACM
  Computing Surveys}\ }\textbf {\bibinfo {volume} {21(3)}},\ \bibinfo {pages}
  {383} (\bibinfo {year} {1989})}\BibitemShut {NoStop}%
\bibitem [{\citenamefont {Scemama}\ \emph {et~al.}(2015)\citenamefont
  {Scemama}, \citenamefont {Giner}, \citenamefont {Applencourt}, \citenamefont
  {David},\ and\ \citenamefont {Caffarel}}]{quantum_package}%
  \BibitemOpen
  \bibfield  {author} {\bibinfo {author} {\bibfnamefont {A.}~\bibnamefont
  {Scemama}}, \bibinfo {author} {\bibfnamefont {E.}~\bibnamefont {Giner}},
  \bibinfo {author} {\bibfnamefont {T.}~\bibnamefont {Applencourt}}, \bibinfo
  {author} {\bibfnamefont {G.}~\bibnamefont {David}}, \ and\ \bibinfo {author}
  {\bibfnamefont {M.}~\bibnamefont {Caffarel}},\ }\href {\doibase
  10.5281/zenodo.30624} {\enquote {\bibinfo {title} {Quantum package v0.6},}\ }
  (\bibinfo {year} {2015}),\ \bibinfo {note}
  {doi:10.5281/zenodo.30624}\BibitemShut {NoStop}%
\bibitem [{\citenamefont {Balabanov}\ and\ \citenamefont
  {Peterson}(2005)}]{Balabanov_2005}%
  \BibitemOpen
  \bibfield  {author} {\bibinfo {author} {\bibfnamefont {N.~B.}\ \bibnamefont
  {Balabanov}}\ and\ \bibinfo {author} {\bibfnamefont {K.~A.}\ \bibnamefont
  {Peterson}},\ }\href {\doibase 10.1063/1.1998907} {\bibfield  {journal}
  {\bibinfo  {journal} {J. Chem. Phys.}\ }\textbf {\bibinfo {volume} {123}},\
  \bibinfo {pages} {064107} (\bibinfo {year} {2005})}\BibitemShut {NoStop}%
\bibitem [{\citenamefont {Schmidt}\ \emph {et~al.}(1993)\citenamefont
  {Schmidt}, \citenamefont {Baldridge}, \citenamefont {Boatz}, \citenamefont
  {Elbert}, \citenamefont {Gordon}, \citenamefont {Jensen}, \citenamefont
  {Koseki}, \citenamefont {Matsunaga}, \citenamefont {Nguyen}, \citenamefont
  {Su}, \citenamefont {Windus}, \citenamefont {Dupuis},\ and\ \citenamefont
  {Montgomery}}]{gamess}%
  \BibitemOpen
  \bibfield  {author} {\bibinfo {author} {\bibfnamefont {M.~W.}\ \bibnamefont
  {Schmidt}}, \bibinfo {author} {\bibfnamefont {K.~K.}\ \bibnamefont
  {Baldridge}}, \bibinfo {author} {\bibfnamefont {J.~A.}\ \bibnamefont
  {Boatz}}, \bibinfo {author} {\bibfnamefont {S.~T.}\ \bibnamefont {Elbert}},
  \bibinfo {author} {\bibfnamefont {M.~S.}\ \bibnamefont {Gordon}}, \bibinfo
  {author} {\bibfnamefont {J.~H.}\ \bibnamefont {Jensen}}, \bibinfo {author}
  {\bibfnamefont {S.}~\bibnamefont {Koseki}}, \bibinfo {author} {\bibfnamefont
  {N.}~\bibnamefont {Matsunaga}}, \bibinfo {author} {\bibfnamefont {K.~A.}\
  \bibnamefont {Nguyen}}, \bibinfo {author} {\bibfnamefont {S.}~\bibnamefont
  {Su}}, \bibinfo {author} {\bibfnamefont {T.~L.}\ \bibnamefont {Windus}},
  \bibinfo {author} {\bibfnamefont {M.}~\bibnamefont {Dupuis}}, \ and\ \bibinfo
  {author} {\bibfnamefont {J.~A.}\ \bibnamefont {Montgomery}},\ }\href
  {\doibase 10.1002/jcc.540141112} {\bibfield  {journal} {\bibinfo  {journal}
  {J. of Comput. Chem.}\ }\textbf {\bibinfo {volume} {14}},\ \bibinfo {pages}
  {1347} (\bibinfo {year} {1993})}\BibitemShut {NoStop}%
\end{thebibliography}
\end{document}